\newcommand{\ISPASS} 

\ifdefined\ISPASS
    \documentclass[conference,10pt]{IEEEtran}

\usepackage{booktabs} 

\usepackage{mathptmx} 

\usepackage{fancyhdr}
\usepackage[normalem]{ulem}
\usepackage[hyphens]{url}
\usepackage[sort,nocompress]{cite}
\usepackage[final]{microtype}
\usepackage{amsmath,amssymb,amsfonts}
\usepackage{textcomp}
\usepackage{color}
\usepackage{xcolor}
\usepackage{soul}
\usepackage{multicol}
\usepackage{multirow}
\usepackage{graphicx}
\usepackage{caption}
\usepackage{subcaption}
\usepackage{enumitem}
\usepackage{colortbl}
\usepackage[letterpaper,left=0.75in,right=0.75in,top=1.00in,bottom=1.00in]{geometry}

\usepackage{algorithm}
\usepackage[noend]{algpseudocode}
\usepackage{varwidth}

\usepackage{xspace}
\usepackage{bm}

\usepackage{mathtools}
\usepackage{flushend}

\usepackage{booktabs}

\usepackage[bookmarks=true,breaklinks=true,letterpaper=true,colorlinks,linkcolor=black,citecolor=blue,urlcolor=black]{hyperref}

\def\BibTeX{{\rm B\kern-.05em{\sc i\kern-.025em b}\kern-.08em
    T\kern-.1667em\lower.7ex\hbox{E}\kern-.125emX}}

\usepackage[font=footnotesize,labelfont=bf]{caption}

\DeclarePairedDelimiter\floor{\lfloor}{\rfloor}

\definecolor{green}{rgb}{0.0,0.4,0.0}
\definecolor{orange}{rgb}{0.7,0.4,0.0}

\newcommand{\ignore}[1]{}

\newcommand{\func}[2]{\ensuremath{\operatorname{#1}(#2)}}
\newcommand{\funcref}[1]{\ensuremath{\operatorname{#1}}}

\newcommand{\okay}{\textcolor{black}{$-$ }}

\newcommand{\close}[1]{\textit{\textcolor{gray}{#1}}}

\newcommand{\mline}[1]{\begin{tabular}{@{}l@{}}#1\end{tabular}}

\IEEEoverridecommandlockouts

\title{Specializing Coherence, Consistency, and Push/Pull for GPU Graph Analytics}

\author{Giordano~Salvador$^{\dagger}$, Wesley~H.~Darvin$^{\dagger}$, Muhammad~Huzaifa$^{\dagger}$, \\
Johnathan~Alsop$^{\ddagger}$, Matthew~D.~Sinclair$^{\ast}$, Sarita~V.~Adve$^{\dagger}$\\
$^{\dagger}$University of Illinios at Urbana-Champaign \quad\quad $^{\ddagger}$AMD Research \quad\quad $^{\ast}$University of Wisconsin-Madison \thanks{This work was supported in part by the Applications Driving Architectures (ADA) Research Center, a JUMP Center co-sponsored  by  SRC  and  DARPA, a Google Faculty Research Award, and the National Science Foundation under grant CCF 16-19245.}
}

\begin{document}

\maketitle

\pagestyle{plain}

\begin{abstract}

This work provides the first study to explore the interaction of update propagation with and without fine-grained synchronization (push vs.\ pull), emerging coherence protocols (GPU vs.\ DeNovo coherence), and software-centric consistency models (DRF0, DRF1, and DRFrlx) for graph workloads on emerging integrated GPU-CPU systems with native unified shared memory. We study 6 graph applications with 6 graph inputs for a total of 36 workloads running on 12 system (hardware+software) configurations reflecting the above design space of update propagation, coherence, and memory consistency. We make three key contributions. First, we show that there is no single best system configuration for all workloads, motivating systems with flexible coherence and consistency support. Second, we develop a model to accurately predict the best system configuration -- this model can be used by software designers to decide on push vs.\ pull and the consistency model and by flexible hardware to invoke the appropriate coherence and consistency configuration for the given workload. Third, we show that the design dimensions explored here are inter-dependent, reinforcing the need for software-hardware co-design in the above design dimensions. For example, software designers deciding on push vs.\ pull must consider the consistency model supported by hardware -- in some cases, push maybe better if hardware supports DRFrlx while pull may be better if hardware does not support DRFrlx.

\end{abstract}

\section{Introduction}
\label{sec:introduction}

As technology scaling 
slows down, architects are increasingly relying on alternative designs that can provide continued performance scaling with the available transistors~\cite{Eeckhout2017}.
\textit{Specialization} is one such approach where a processing element is designed to perform a specific task.
These specialized processors make more efficient use of transistors in terms of area and performance, but are harder to program and less widely applicable than general purpose processors.
To get around these limitations, system designers also leverage \textit{heterogeneity} of processing elements by providing many different kinds of processing elements, each suited to different uses.
The graphics processing unit~(GPU) is an example of a specialized processor which is both being integrated on-chip with central processing units~(CPUs) and becoming more general purpose.
This allows application developers to accelerate parallel portions of their applications with a GPU architecture that is designed for large-scale parallelism and memory throughput.

The domain of graph analytics is one that has been shown to benefit from increasingly specialized processors and software stacks.
Many frameworks exist that generalize and provide abstractions for programming and optimizing graph workloads~\cite{NguyenLenharth2013,WangPan2017,ShunBlelloch2013,BurtscherNasre2012,GrossmanLitz2018,SundaramSatish2015,RoyMihailovic2013,ZhangYang2018,HongChafi2012,ZhangChen2015,KyrolaBlelloch2012,KhorasaniVora2014,SabetQiu2018}.
While numerous frameworks focused on distributed and shared memory multi-core CPU systems, some have been implemented for the GPU with the goal of providing better performance and scalability than the more traditional CPU graph frameworks.
\textit{Topology-driven}, \textit{vertex-centric} implementations~\cite{CheBoyer2009,KimBatten2014,KhorasaniVora2014} of graph algorithms are commonly used because they are easy to implement and can achieve high throughput on GPUs via their asynchronous thread structure.
A significant algorithmic design choice affecting these implementations is that of {\em update propagation} in terms of \textit{push} vs.\ \textit{pull}~\cite{BestaPodstawski2017}.
For kernels of graph algorithms that need to perform edge-propagated updates, a source vertex can either pull changes \emph{from} its neighbors or push changes \emph{to} its neighbors.
For a given algorithm, both choices preserve the direction of the edge-propagated updates through the edge, but can change the control-flow and memory access behavior of the implementation.
More importantly, pull does not require fine-grained synchronization while push does.

Traditional GPU architectures suffer from unique challenges which cause fine-grained synchronization to be performance prohibitive.
Recently, however, with CPU-GPU integration and native support for a global (shared) address space, new coherence protocols and software-centric memory consistency models have been proposed and/or commercially implemented which make fine-grained synchronization more efficient on GPUs.
For example, in contrast to conventional GPU coherence protocols (referred to as GPU coherence~\cite{SinclairAlsop2015}), the DeNovo protocol~\cite{SinclairAlsop2015} uses ownership to exploit synchronization locality in L1 caches.
Similarly, in contrast to the simple DRF0 consistency model, the DRF1 model avoids heavyweight invalidation and dirty data flushing at unpaired synchronization points and the DRFrlx model further allows reordering relaxed synchronization operations~\cite{SinclairAlsop2017}.

This work provides the first study that explores the interaction of update propagation with and without fine-grained synchronization (push vs.\ pull), emerging coherence protocols (GPU vs.\ DeNovo coherence), and software-centric consistency models (DRF0, DRF1, and DRFrlx) for graph workloads on emerging integrated GPU-CPU systems with native unified shared memory. The paper makes the following contributions.

    {\bf Need for flexibility -- one size does not fit all:} We evaluate the impact of graph algorithm and graph input on the hardware+software design decisions for update-propagation (push or pull), coherence protocol (GPU or DeNovo), and consistency model (DRF0, DRF1, or DRFrlx). We study 6 graph algorithms and 6 graph inputs for a total of 36 workloads on the 12 possible configurations of hardware+software (referred to as system below). We find that there is no one configuration that is optimal for all cases. The configuration with push, GPU coherence, and DRFrlx consistency model performs the best for 24 (majority) of the workloads, but for the remaining 12, different configurations perform significantly better (reducing execution time by up to 87\%, average of 44\%). This motivates systems with flexible consistency models and flexible coherence protocols such as Spandex~\cite{AlsopSinclair2018} which can adapt to the different needs of different graph workloads.
    
    
   {\bf A model to predict the best design choice:} We develop a simple model to predict the best performing system configuration (of the 12 choices for update propagation, coherence, and consistency combinations) for a given graph algorithm and input.
     The model uses only 6 simple parameters to represent the graph algorithm and input, which drive a decision tree to choose the optimal system configuration.
    %
    For each of the 36 workloads, the model picks a configuration that performs within $3.5$\% of the best one. For 28 of the 36 cases, it picks exactly the optimal configuration.
    This model can be used by software developers to determine the appropriate software configuration and by system designers to drive configuration selection for an adaptive system such as Spandex~\cite{AlsopSinclair2018}.
    
    {\bf Inter-dependent design dimensions:} Our evaluation also shows that the three design dimensions (update propagation, coherence, and consistency) are interdependent; choosing one without consideration of the support for others can lead to significantly sub-optimal designs. For example, the choice of push vs. pull depends, in general, on whether the system supports DRFrlx. Specifically, for graphs with high load imbalance, assuming the system supports DRFrlx leads to a choice of push. However, running push on a system that supports only DRF1 instead of DRFrlx can perform up to 80\% worse than a pull based implementation. Thus, design decisions for pull vs.\ push that ignore consistency support can lead to significantly sub-optimal decisions. 
    Recognizing such interdependences, we also show how to extend our model to determine the optimal configuration assuming only partial support for the design space considered.
    

{\bf Summary:} Overall, this work is the first to systematically consider the impact of emerging coherence and consistency options on graph algorithms running on integrated GPUs. Our results motivate support for flexible coherence and consistency, we show a model to predict the best configuration for a graph algorithm + input and to drive flexible systems, and we demonstrate the interdependence of the design decisions of push vs.\ pull, consistency, and coherence supported.

\section{Design Space}
\label{sec:comm-design-space}

\newcommand{\PullI}{\mline{Pull}}
\newcommand{\PullD}{\mline{Target in outer loop;\\}}
\newcommand{\PullF}{\mline{Dense local updates;\\Sparse remote reads;\\Elide work at sources;\\}}

\newcommand{\PushI}{\mline{Push}}
\newcommand{\PushD}{\mline{Source in outer loop;\\}}
\newcommand{\PushF}{\mline{Dense local reads;\\Sparse remote atomics;\\Elide work at targets;\\}}

\newcommand{\PpI}{\mline{Push+Pull}}
\newcommand{\PpD}{\mline{Non-deterministic\\source/target direction;\\}}
\newcommand{\PpF}{\mline{Remote reads and updates;\\}}

\newcommand{\GpuI}{\mline{GPU}}
\newcommand{\GpuD}{\mline{Write-through at L1 on sync.;\\Self-inval. at L1 on sync.;\\}}
\newcommand{\GpuF}{\mline{Atomics at L2~(bypass L1);\\Good when low update reuse;\\}}

\newcommand{\DnvI}{\mline{DeNovo}}
\newcommand{\DnvD}{\mline{Registration~(ownership) at L1\\on sync.;\\}}
\newcommand{\DnvF}{\mline{Atomics at L1;\\Good when high update reuse;\\}}

\newcommand{\ZeroI}{\mline{DRF0}}
\newcommand{\ZeroD}{\mline{Data to data reordering;\\Seq. consistency for\\acquires\&releases;\\}}
\newcommand{\ZeroF}{\mline{Overlapping data accesses\\w/ data updates;\\Programmability;\\}}

\newcommand{\OneI}{\mline{DRF1}}
\newcommand{\OneD}{\mline{Data to data reordering;\\Data to atomics reordering;\\}}
\newcommand{\OneF}{\mline{Overlapping data accesses w/\\atomics;\\Programmability;\\}}

\newcommand{\RlxI}{\mline{DRFrlx}}
\newcommand{\RlxD}{\mline{Data-data reordering;\\Data-atomic reordering;\\Atomic-atomic reordering;\\}}
\newcommand{\RlxF}{\mline{Overlapping data accesses w/\\atomics;\\Overlapping atomics;\\Mitigate imbalance via MLP;\\}}

\begin{table}[!t]
    \centering
    \captionsetup{font=footnotesize}
    \caption{Summary of the implementation design space and the salient features explored by the model described in this work.
    Salient features are introduced in Section~\ref{sec:comm-design-space} and then discussed in the context of specialization in Section~\ref{sec:model}.}%
    \scriptsize
    
    \begin{tabular}{c | l l  }
        \textbf{Implementation}    & \textbf{Description}   & \textbf{Salient Features}     \\
    \toprule
        \multicolumn{3}{c}{Push vs.\ Pull Dimension}\\
        \hline
        \PullI            & \PullD        & \PullF      \\
        \arrayrulecolor{gray}
       \hline
        \PushI            & \PushD        & \PushF      \\
       \hline
        \PpI              & \PpD          & \PpF        \\
        \arrayrulecolor{black} 
     \toprule 
        \multicolumn{3}{c}{Coherence Dimension}\\
        \hline
        \arrayrulecolor{gray}
        \GpuI             & \GpuD         & \GpuF       \\
        \hline
        \DnvI             & \DnvD         & \DnvF       \\
        \arrayrulecolor{black} 
    \toprule              
        \multicolumn{3}{c}{Consistency Dimension}\\
        \hline
        \arrayrulecolor{gray}
        \ZeroI            & \ZeroD        & \ZeroF      \\
        \hline
        \OneI             & \OneD         & \OneF       \\
        \hline
        \RlxI             & \RlxD         & \RlxF       \\
    \arrayrulecolor{black} 
    \bottomrule
    \end{tabular}%

    \normalsize
    \label{tbl:implementation}
\end{table}

\begin{figure*}[!t]
  \small
  \hrulefill
  \begin{multicols}{2}
  \begin{algorithmic}[1]
    \Procedure{PushKernel}{$ V, \funcref{E_{in}}, \funcref{E_{out}}, \funcref{op}, \funcref{vprop}, i $}
      \For{$ s \in V $}
        \If{$ \func{spred}{s} $}                                                                   
          \State $ p_s \gets \func{vprop}{s, i} $
          \For{$ e \in \func{E_{out}}{s} $}
            \State $ t \gets \func{dest}{e} $ 
            \If{$ \func{tpred}{t} $}                                                               
              \State 
                $ \func{atomicUpdate}{\funcref{op}, t, p_s, i + 1} $
            \EndIf
          \EndFor
        \EndIf
      \EndFor
      \State
      \State
    \EndProcedure
  \end{algorithmic}\par
  \begin{algorithmic}[1]
    \Procedure{PullKernel}{$ V, \funcref{E_{in}}, \funcref{E_{out}}, \funcref{op}, \funcref{vprop}, i $}
      \For{$ t \in V $}
        \If{$ \func{tpred}{t} $}                                                                   
          \State $ p_t \gets \func{init}{} $
          \For{$ e \in \func{E_{in}}{t} $}
            \State $ s \gets \func{src}{e} $ 
            \If{$ \func{spred}{s} $}                                                               
              \State $ p_s \gets \func{vprop}{s, i} $
              \State $ p_t \gets \func{op}{p_t, p_s} $                        
            \EndIf
          \EndFor
          \State $ \func{update}{t, p_t, i + 1} $
        \EndIf
      \EndFor
    \EndProcedure
  \end{algorithmic}%
  \end{multicols}%
  \vspace{-0.25in}
  \hrulefill
  \captionsetup{font=footnotesize}
  \caption{Vertex-centric push and pull kernels with static algorithmic traversal. The $\operatorname{spred}$ and $\operatorname{tpred}$ functions influence the algorithmic control of the workload. The $\operatorname{vprop}$, $\operatorname{update}$, and $\operatorname{atomicUpdate}$ functions influence the algorithmic information of the workload. The iteration variable $i$ is used to track the state of the double-buffered vertex properties between kernel invocations, such that the previous iteration's values are read-only.}%
  \label{alg:push-pull}%
\vspace{-.25in}
\end{figure*}%


To study the interaction of irregular workloads with the underlying communication fabric, we first describe an implementation design space comprised of three dimensions: push vs. pull, cache coherence protocols, and memory consistency models.
Table~\ref{tbl:implementation} summarizes this design space and the salient trade-off considerations for each implementation.
Each implementation represents a design specialization that can be made for the irregular graph workloads that we explore.
The salient features of these implementations are used by our \textit{specialization model} in~Section~\ref{sec:model}.

  \subsection{Push vs. Pull}
  \label{subsec:space-pvp}

  We define push vs. pull as the dimension where updates may be to shared remote vertex properties or unshared local vertex properties, respectively.
  
  Figure~\ref{alg:push-pull} shows pseudocode for two representative kernels: one using push updates and another using pull updates.
  Interest in this dimension is motivated by considerations for fine-grained synchronization overhead, as well as the elision of redundant work~(described in depth in Sections~\ref{sec:taxonomy}).
  
  Although push updates incur additional overheads via the use of atomics, optimizations for atomics in the coherence and consistency dimensions are available to maintain or even improve performance.
  Furthermore, as shown by line 2 in Figure~\ref{alg:push-pull}, dense traversal of source nodes in the graph at the outer loop reduces the total number of inner-loop accesses to the source node.
  
  Pull updates do not require atomics, as each vertex is only updated by its assigned thread.
  As seen in lines 5--9 of Figure~\ref{alg:push-pull}, pull implementations must wait on sparse remote reads before performing their local updates~(line 10) because they traverse source nodes in the graph sparsely~(line 5).
  However, dense local updates and traversing each target node once in the outer loop can be beneficial, depending on the workload's properties.
  
  In some scenarios, both push and pull can be employed together, requiring both reads and updates to use synchronization.
  This scenario presents a third implementation option of push+pull, which is dependent upon the construction of the algorithm.
  

  \subsection{Coherence}
  \label{subsec:space-coherence}

  GPUs typically use simple, software-driven coherence protocols that rely on data-race-freedom and assume synchronization happens infrequently and at a coarse granularity. We consider a commonly used protocol, referred to as GPU coherence, as follows~\cite{SinclairAlsop2015}.
  At synchronization reads, GPU coherence self-invalidates the entire L1 cache, and at synchronization writes, it invokes write throughs (flushes) of all dirty data to the shared last level cache (LLC).
  All synchronization (also referred to as atomic) accesses execute at the LLC and in program order (normal data accesses are still performed at the L1).
  This coherence protocol enables sequentially consistent (SC) execution for any program that is data-race-free, also referred to as the DRF0 consistency model and discussed below.\footnote{Recent systems provide for the heterogeneous-race-free (HRF)~\cite{HowerHechtman2014} consistency model which adds lower-overhead locally scoped synchronization. However, later work~\cite{SinclairAlsop2015,AlsopOrr2016} showed that HRF comes at an unnecessary programmability cost since coherence protocols such as DeNovo~\cite{SinclairAlsop2015} and the hLRC variant~\cite{AlsopOrr2016} support DRF with similar (and sometimes better) efficiency than GPU coherence+HRF and with easier programmability. We therefore do not consider HRF here. We also note that DRFrlx provides similar performance benefits as HRF with GPU coherence, thereby providing representative coverage of the coherence+consistency design space.}

  While GPU coherence is simple and performs well for traditional GPU workloads, it can be inefficient for workloads with frequent synchronization.
  The DeNovo coherence protocol seeks to address these inefficiencies and has been shown to provide high performance for a wide variety of applications in heterogeneous systems~\cite{SinclairAlsop2015, AlsopSinclair2018}.
  DeNovo obtains exclusive ownership for written data and atomic accesses. Owned data does not have to be invalidated or flushed at synchronization, and synchronization/atomics for owned locations can execute at the local L1. 
  This significantly reduces the overhead of synchronization accesses and improves reuse of written data and atomics, while still avoiding the overheads of writer-initiated invalidations (as present in standard CPU MESI-style protocols).
  
  However, for workloads where L1 cache utilization or reuse is poor, DeNovo can add overhead relative to GPU coherence by unnecessarily requiring ownership registration and unnecessarily bringing in data for atomic updates into the L1 and replacing other useful data. Thus, there is a workload dependent tradeoff between GPU coherence and DeNovo.
  In the design dimension of coherence protocols, we consider both GPU and DeNovo coherence.

  \subsection{Consistency}
  \label{subsec:space-consistency}
  
  Most programming languages (and GPUs) support the data-race-free family of consistency models. DRF0 (data-race-free 0) is the simplest to program of these models, providing sequential consistency (SC) to data-race-free programs. With GPUs, DRF0 incurs data invalidation and flush overhead on all synchronization accesses as described above (only for non-owned data with DeNovo). To partly mitigate this overhead, DRF1 recognizes that some synchronization accesses (distinguished as unpaired synchronization) are never used to order data accesses, and so can be overlapped (reordered) with data accesses~\cite{AdveHill1993,SinclairAlsop2017}. For such synchronizations, the only overhead is that they need to appear SC relative to other synchronization accesses; therefore, they are executed in program order relative to each other (and at the LLC for GPU coherence). The programmer still gets SC for DRF programs, but the definition of a DRF program is enhanced to accommodate unpaired synchronizations. Thus, DRF1 offers better performance than DRF0 (by allowing reordering of data accesses relative to unpaired synchronizations), but with slightly more complex programmability.
  
  
  
  Finally, data-race-free-relaxed (DRFrlx)~\cite{SinclairAlsop2017} recognizes that some synchronizations do not need to be ordered even with respect to each other. Thus, DRFrlx potentially improves performance by exploiting intra-thread memory level parallelism (MLP) among such relaxed synchronizations (popularly known as relaxed atomics). While several GPUs support relaxed atomics, they are known to be hard to reason about and not all systems provide efficient support for them~\cite{SinclairAlsop2015}.

  
  In the memory consistency models design dimension, we study DRF0, DRF1 and DRFrlx.

\section{Taxonomy for Graph Analytics}
\label{sec:taxonomy}

In this section, we describe workload properties that determine the computation performed by each workload, and subsequently, the required communication.
These properties form the taxonomy used by our work in the context of traditional vertex-centric graph analytics applications.
For our taxonomy, we first define each \textit{workload} as an input graph and algorithm pair.
We then describe two salient features of graph workloads affecting communication: \textit{graph structure}, and \textit{algorithmic properties}.

  \subsection{Graph Structure}
  \label{subsec:taxonomy-gs}

  Graph structure is the connectivity of the input graph
  and influences both the working-set size and the data access patterns of a workload.
  In addition to the effect of algorithmic properties, the overall amount of communication performed by a workload depends on graph size, vertex connectivity, and the degree distribution across vertices.
  Thus, we highlight three primary components of graph structure that influence communication: \emph{volume}, \emph{reuse}, and \emph{imbalance}.
  %


    \subsubsection{Volume}
    \label{subsubsec:taxonomy-volume}
  
    The volume of data that is accessed is considered the working set size of the application, and is thus a function of the vertices and edges in the graph.
    For some algorithms, auxiliary structures which scale with the number of vertices and edges add to the size of the working set (e.g.,\ a color array for graph coloring).
    Although volume can also depend on the frontier size of a given iteration, the frontier size is dependent on dynamic behavior of the workload, which is difficult to compute statically.
    
    As a \emph{proxy} for the working set size of a core, we compute the volume per core as an average distribution of the graph over all cores:
    \begin{equation}
        Volume(G) =
        \frac{|V(G)| + |E(G)|}{|SM|} \,.
        \label{eqn:vol}
    \end{equation}
    In Equation~\ref{eqn:vol}, $G$ is the input graph, $V$ and $E$ are the vertex and edge sets respectively, and $|SM|$ is the number of GPU cores available.
    The $Volume$ metric thus gives a value proportional to the average working set size to be touched by each GPU core.
    To anticipate whether this volume is high or low, we compare the volume of data to the available L1 and L2 resources that can be dedicated to each core during full utilization of the GPU.
    If the volume can roughly fit in each core's dedicated L1 data cache, then the volume is low.
    If the volume can fit within a core's allocation of the L2 cache~(assuming an even partitioning across cores), then the volume is medium.
    Otherwise, the volume is high.
    
    These three discretized values of volume act as inputs to our specialization model~(Section~\ref{sec:model}).

      \subsubsection{Reuse}
      \label{subsubsec:taxonomy-reuse}

\begin{figure}[!t]
  \centering
  \begin{subfigure}[t]{0.5\linewidth}
    \centering
    \includegraphics[keepaspectratio,width=\linewidth]{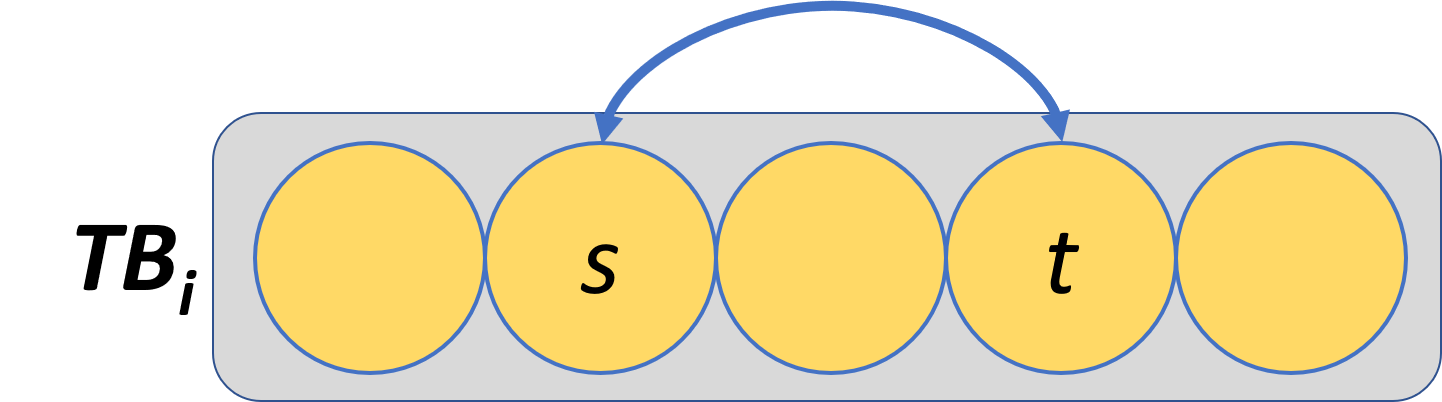}
    \caption{Local Neighbors~($AN_L$)}
    \label{subfig:anl}
  \end{subfigure}
  \begin{subfigure}[t]{0.5\linewidth}
    \centering
    \includegraphics[keepaspectratio,width=\linewidth]{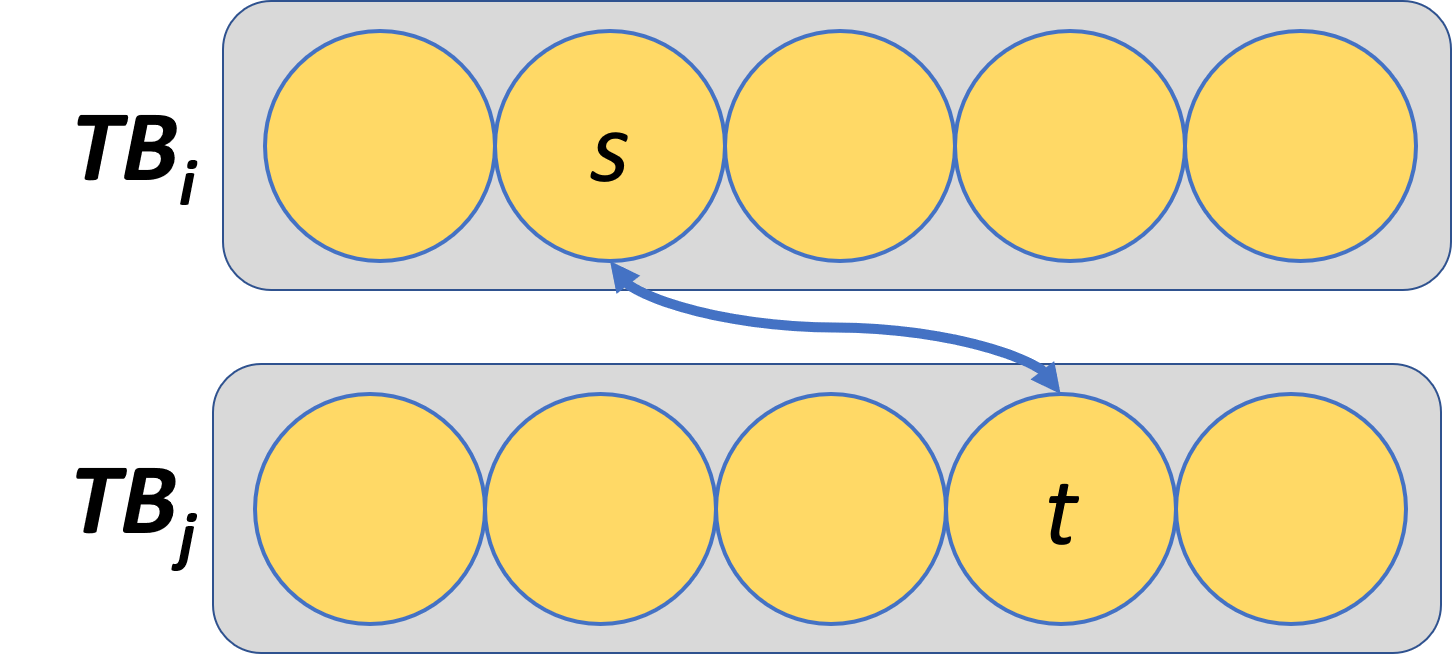}
    \caption{Remote Neighbors~($AN_R$)}
    \label{subfig:anr}
  \end{subfigure}
  \captionsetup{font=footnotesize}
  \caption{Depictions of various locality scenarios involving threads within and across thread blocks.}%
  \label{fig:tax-locality}%
\vspace{-.25in}
\end{figure}%

      \begin{figure}[!t]
    \centering
    \setlength{\abovedisplayskip}{4pt}
    \setlength{\belowdisplayskip}{4pt}
    \begin{align}
        TB_L(v_1, v_2) =  
            \begin{cases}
                0 & v_1=v_2 \\
				1 & \floor{v_1/|TB|}=\floor{v_2/|TB|}\\
                0 & otherwise
            \end{cases}%
        \label{eqn:tb-l}%
    \end{align}%
    \setlength{\belowdisplayskip}{4pt}
    \setlength{\abovedisplayskip}{4pt}
    \begin{align}
        TB_R(v_1, v_2) =  
            \begin{cases}
                0 & v_1=v_2 \\
                1 & \floor{v_1/|TB|} \neq \floor{v_2/|TB|}\\
                0 & otherwise
            \end{cases}%
        \label{eqn:tb-r}%
    \end{align}%
    \setlength{\belowdisplayskip}{4pt}
    \setlength{\abovedisplayskip}{4pt}
    \begin{align}
            AN_L(G) =  
            \left( \sum\limits_v^{V(G)} \sum\limits_n^{E(v)} TB_L(v,n) \right) \frac{1}{|V(G)|}
        \label{eqn:an-l}%
    \end{align}%
    \setlength{\belowdisplayskip}{4pt}
    \setlength{\abovedisplayskip}{4pt}
    \begin{align}
            AN_R(G) = 
            \left( \sum\limits_v^{V(G)} \sum\limits_n^{E(v)} TB_R(v,n) \right)  \frac{1}{|V(G)|}
        \label{eqn:an-r}%
    \end{align}%
    \setlength{\belowdisplayskip}{4pt}
    \setlength{\abovedisplayskip}{4pt}
    \begin{align}
            Reuse(G) =
            \frac{1}{2} \left( 1 + \frac{AN_L(G) - AN_R(G)}{|E(G)| / |V(G)|} \right)
        \label{eqn:loc}
    \end{align}%
    \setlength{\belowdisplayskip}{4pt}
    \normalsize
    \captionsetup{font=footnotesize}
    \caption{Heuristics for quantifying the reuse scenarios illustrated in Figure~\ref{fig:tax-locality}.
             In these equations, $G$ is the input graph, $V$ is the set of vertices, $E$ is the set of edges, and $|TB|$ is the thread block size.
             Average number of local neighbors~($AN_L$) and average number of remote neighbors~($AN_R$) are computed using immediate neighbors~($n$) of the source vertex~($v$).
             The helper functions, thread block local~($TB_L$) and thread block remote~($TB_R$), are used to compute whether two vertices~($v_1$ and $v_2$) share a thread block.
             Finally, the $Reuse$ function computes a locality metric based on $AN_L$, $AN_R$, and the average degree of $G$.
            }%
    \label{fig:equations}%
\vspace{-0.25in}
\end{figure}%


    %

      To drive our specialization model, we need to anticipate how much intra-thread-block data reuse can be exploited in our graph workloads based on vertex connectivity.
      Figure~\ref{fig:tax-locality} depicts connectivity attributes we use that describe the potential for reuse.
      Our heuristics assume directed, symmetric graphs to match the universal input format we have for our workloads; however, they could be extended to express reuse for directed, non-symmetric graphs.
      
      Figure~\ref{subfig:anl} depicts the simplest case where neighbors in the graph are in the same thread block~(\textit{local neighbors}, see Equation~\ref{eqn:an-l}).
      Such vertices are scheduled concurrently~(by thread block), and may exploit data reuse on reads and updates to shared structures.
      
      On the other hand, Figure~\ref{subfig:anr} depicts the scenario where neighbors are placed in different thread blocks~(\textit{remote neighbors}, see Equation~\ref{eqn:an-r}).
      Achieving reuse for updates through these edges would be less likely and heavily schedule-dependent.
      We introduce the $Reuse$ metric~(Equation~\ref{eqn:loc}), combining Equations~\ref{eqn:an-l} and~\ref{eqn:an-r}.
      Observe that the $AN_L$ and $AN_R$ heuristics sum to the average degree of the graph, each considering the disjoint subsets of local edges and remote edges, respectively.
      We subtract these values and normalize to the average degree of the whole graph, resulting in a term~(seen in Equation~\ref{eqn:loc}) that ranges from -1 to 1 that shows how heavily edges are skewed towards remote connectivity and local connectivity, respectively.
      As a last step, we transform the range of the final value of reuse to a more intuitive range of 0 to 1, where 0 indicates low potential for reuse and 1 indicates high potential for reuse. Our specialization model~(Section~\ref{sec:model}) takes as input this computed value of reuse.

      \subsubsection{Imbalance}
      \label{subsubsec:taxonomy-imbalance}
      
      Imbalance of a workload is affected by the degree distribution of the graph.
      Disproportionate work allocation across the warps of a thread block can cause scenarios where most of the block's warps have completed, but a small number of warps continue to process large degree vertices.
      Since each GPU core has a limited number of thread blocks that can be scheduled, long running warps in these imbalanced thread blocks limit the pool of warps able to be scheduled for a core.
      
      To compute the magnitude of imbalance, we classify warps in a thread block which contribute to high imbalance in that thread block.
      This classification is done by clustering warps in a thread block by the max degree processed by those warps.
      We accomplish this clustering via the k-means clustering algorithm, such that we search the warp for two clusters: low max degree and high max degree.
      If the difference of the final centroids' values is larger than some threshold, the thread block is considered imbalanced.
      The graph balance metric is computed to be the ratio of imbalanced thread blocks to total thread blocks~(a value from 0 to 1, inclusive).
      The larger this value, the more the imbalance is present in the graph.
      Thus, imbalance is measured as follows:
      \begin{equation}
          Imbalance(G) =
          \frac{1}{|V(G)| / |TB|} \left( \sum\limits_{tb}^{TB(G)} marked(tb) \right) \,.
          \label{eqn:imb}
      \end{equation}
      In Equation~\ref{eqn:imb}, $G$ is the input graph, $V$ the set of vertices, $TB$ is the set of thread blocks, and $marked$ is a binary function which evaluates to 1 when
      the given thread block is imbalanced, and 0 otherwise.
  
      We make no assumptions about the ordering of our graphs (i.e.,\ we do not assume a highest degree vertex ordering) so this computation must be performed for all warps.
      The computed value of imbalance is provided as an input to our specialization model~(Section~\ref{sec:model}).

  \subsection{Algorithmic Properties}
  \label{subsec:taxonomy-ap}

  We analyze these properties of graph applications to anticipate the algorithmic influence on data access patterns, and thus use them for specialization of communication.
  We identify three salient components of algorithmic properties in this regard: \textit{algorithmic traversal}, \textit{algorithmic control}, and \textit{algorithmic information}.
  After introducing these components in this section, we describe how we model them for use for specialization in Section~\ref{sec:model}.
  
  
    \subsubsection{Algorithmic Traversal}
    \label{subsec:taxonomy-at}
  
    Algorithmic Traversal is the pattern in which each work item propagates changes through the input graph; i.e., algorithmic traversal describes \emph{where} in the graph information is propagated.
    An algorithm's traversal can be either \textit{static} or \textit{dynamic}.
  
    A static traversal is one in which the connectivity of the input graph is directly used to define the source and target nodes.
    That is, in a static traversal, the source and target nodes of an update are always neighbors in the input graph.
  
    In algorithms with a dynamic traversal, the source and target nodes are data-dependent and dynamically computed such that they may not be neighboring nodes in the input graph.
    An example of dynamic traversal is when updates propagate over the edges defined by the transitive closure of the input graph; i.e., edges that are not present in the source graph).
  
    We analyze each application to determine its traversal type, static or dynamic, and use it in our specialization model~(\ref{sec:model}).
  
    \subsubsection{Algorithmic Control}
    \label{subsec:taxonomy-ac}
  
    Algorithmic control is the control flow that dictates \emph{whether} information is propagated or not.
    For vertex-centric algorithms, control flow is affected when the algorithm necessitates that information should only be propagated for a vertices in the active set of the graph; i.e., the \textit{frontier}~\cite{WangPan2017}.
    Algorithmic control manifests in code as predicates that are functions of properties of the vertices on an edge.
    
    Figure~\ref{alg:push-pull} shows pseudocode for both a push and pull graph kernel, with predicates for source and target vertices denoted by $\operatorname{spred}$ and $\operatorname{tpred}$, respectively.
    The position of each predicate in the control flow changes from the outer loop~(line 3) to the inner loop~(line 7), or vice versa, when the algorithm is changed between push and pull, respectively.
    Therefore, based on whether the implementation is push or pull, and if one predicate is more likely to be true than the other, more updates and their computations can be elided in the outer loop.
    
    Therefore, algorithmic control is specified as \textit{source} if push elides more computations and updates, \textit{target} if pull elides more computations and updates, and \textit{symmetric} if both push and pull elide an equal amount of work. The value that algorithmic control takes on for a particular application is used by our specialization model~(Section~\ref{sec:model}).
  
    Finally, there are some algorithms which perform both racy push and pull updates in the same dynamic traversal~(see Section~\ref{subsec:taxonomy-at}).
    These algorithms do not have a design space which can leverage asymmetry in algorithmic control between push and pull implementations.

    \subsubsection{Algorithmic Information}
    \label{subsec:taxonomy-ai}
  
    Algorithmic information is the data that is used to compute control flow and the updated value of a vertex; i.e., algorithmic information describes \emph{what} information is propagated, and influences the working set of the workload.
    In Figure~\ref{alg:push-pull}, algorithmic information is shown as accesses to source vertex properties through the function $\operatorname{vprop}$, and implicitly for the target vertex properties through the functions $\operatorname{update}$ and $\operatorname{atomicUpdate}$.
    
    As with control, the position of each property access changes from the outer loop to the inner loop, or vice versa, between push and pull algorithms, respectively.
    As a consequence, an algorithm that performs either push or pull may have better expected performance due to hoisting redundant data accesses and computations into the outer loop.
    
    Therefore, algorithmic information is specified as \textit{source} if push hoists more computations and updates, \textit{target} if pull hoists more computations and updates, and \textit{symmetric} if both push and pull hoist an equal amount of work.
    The value that algorithmic information takes on for a particular application is used by our specialization model~(Section~\ref{sec:model}).
  
    Finally, algorithms which perform both racy push and pull updates in the same traversal~(see Section~\ref{subsec:taxonomy-at}) cannot leverage asymmetry in information between push and pull implementations, because both pull and push updates are present throughout the loop body.
\section{Workload-Driven Specialization}
\label{sec:model}

This section introduces a model that establishes how to specialize the update-propagation of the algorithm and the coherence and consistency of the memory system based on the application and input-dependent features introduced in our taxonomy~(see Section~\ref{sec:taxonomy}).


    \subsection{Full Design Space Optimization}
    \label{subsec:model-full}
    
\begin{figure}[!t]
  \centering
    \includegraphics[keepaspectratio,width=\linewidth]{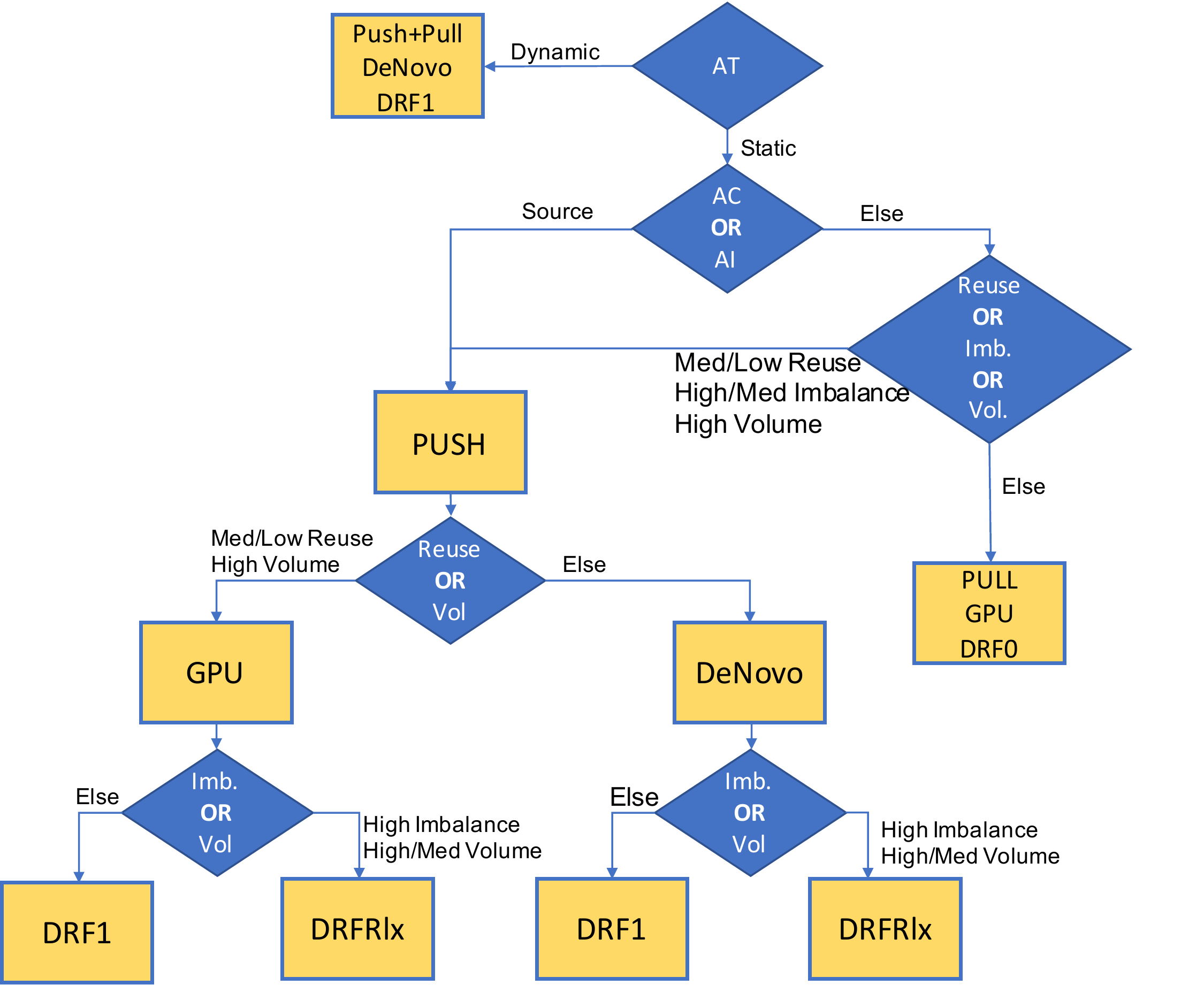}
  \captionsetup{font=footnotesize}
  \caption{Decision tree for specialization when considering our full design space.  Diamonds contain a conditional, paths are labeled with the characterizations that satisfy the conditional for that path.}
  \label{fig:model-flow}%
\vspace{-.25in}
\end{figure}

    This section elaborates on how and why our taxonomy properties described in Sections~\ref{subsec:taxonomy-gs} and~\ref{subsec:taxonomy-ap} interact with each other.
    Figure~\ref{fig:model-flow} contains a decision tree that streamlines these choices resulting in a well optimized configuration.
    
    \subsubsection{Push and Pull}
    \label{subsubsec:model-full-push-pull}
    As stated in Sections~\ref{subsec:taxonomy-ac} and~\ref{subsec:taxonomy-ai}, both algorithmic control and algorithmic information make decisions with respect to actions on either the source or target.
    Eliding work or hoisting loads at source is sufficient for recommending a push algorithm, regardless of what the input graph is.
    However, eliding work or hoisting loads at the target to recommend a pull algorithm is not sufficient.
    This raises the question, why is push more preferred over pull?
    Simply, a push algorithm makes dense reads (on the critical path) and then issues sparse writes (off the critical path).
    In a pull algorithm, there are blocking sparse reads.  These are likely to not perform as well as a push algorithm unless the input has particular properties. 
    
    Even with considering a pull implementation, it is sufficient for the input graph to have medium or low reuse, or medium or high imbalance, or high volume to favor a push implementation instead.
    If the input does not maximize data reuse there is limited benefit to bringing it into the local cache, mitigating one of the main reasons to select pull.
    Similarly, a high volume input will hinder any possible reuse by thrashing the cache.
    A medium or high imbalance value signals that push performance can be improved by allowing data and/or atomic reordering with respect to atomics. This choice (DRF1 vs.\ DRFrlx) will be discussed further in Section~\ref{subsubsec:model-full-consistency}, but is sufficient for us to recommend push.
    
    If the criteria is met to recommend a pull configuration, we pair it with GPU coherence and DRF0.
    The non-atomic reads and updates of a pull implementation interface well with GPU coherence (as described in Section~\ref{subsec:space-consistency}).
    Similarly, non-atomic reads and updates mean there is no need for atomic relaxation and DRF0 will perform just as well.
    
    \subsubsection{Coherence}
    \label{subsubsec:model-full-coherence}
    Assuming a push implementation was selected, the next design space dimension to select is coherence.
    If the graph input has medium or low reuse, or high volume, we recommend GPU coherence.
    Given a push implementation, the key difference between DeNovo and GPU coherence is that DeNovo performs owned (i.e.,\ L1) atomics.
    If there is medium or low reuse, there is little anticipated benefit to bringing data into the L1 for atomics.
    Similarly, no matter how much reuse we observe, if there is high volume we are unlikely to be able to leverage it as we anticipate cache thrashing.
    If neither of these conditions are met, we can expect that DeNovo will have a performance benefit.
    
    \subsubsection{Consistency}
    \label{subsubsec:model-full-consistency}
    Regardless of the coherence protocol selected, we must reason about the optimal consistency model.
    It is sufficient for the input graph to have high imbalance, or high or medium volume to recommend DRFrlx.
    A high imbalance characteristic means that we have several long-running warps that are underutilizing GPU hardware and increasing application run-time.
    Given our imbalance metric (described in Section~\ref{subsubsec:taxonomy-imbalance}) and that at this point in the model-flow we know we are utilizing a push implementation, we know there are atomics proportional to vertex degree that need to be processed.
    Using DRFrlx we can utilize MLP and overlap these atomics, improving the performance of imbalanced threads and increasing the overall performance and hardware utilization of the workload.
    With high or medium volume we anticipate that cache thrashing will increase the number of outstanding atomics, which would be similarly susceptible to the MLP improvements provided by DRFrlx.
    Without either of these characteristics we conservatively recommend DRF1, saving the programmer the effort of reasoning about relaxed atomics.
    
    \subsubsection{Algorithmic Traversal}
    \label{subsubsec:model-full-at}
    The model as described above is predicated on the decision that AT is static. If AT is dynamic, by definition we cannot chose a push or pull implementation as traversal is determined at run-time via synchronization accesses, so instead we describe this as a dynamic push+pull implementation.
    For dynamic traversal workloads, the amount of racy accesses over the lifetime of the workload change over time.
    Some of these algorithms have a traversal which \textit{constricts} the number of possible racing accesses over time.
    In these scenarios, we can anticipate that a reduction in likelihood of racy accesses competing for the same data.
    This paired with a reduction in data volume leads us to anticipate data reuse.
    Therefore, we specialize for reuse by using DeNovo coherence, allowing racy accesses to receive ownership at the L1.
  
    Other dynamic traversal algorithms can exist where the number of racy accesses to the same data does \textit{not constrict}.
    We argue that in such a case DeNovo is still preferable.  As the dataset converges, each atomic access bringing data into the L1 will serve an increasing number of requests from threads in that SM.  We believe that this more than offsets the cost of ping-ponging.
    This is compared to GPU coherence which will serialize requests at the L2.  L2 atomics to the same address must be serialized to account for ordering requirements; there is no distinction made between atomic requests from the same thread or different different threads within a warp as it would be infeasible to include something like threadID in the MSHR~\footnote{Such a distinction is not needed when working with in-order networks, but that solution will not scale as GPUs increase core-counts, and is therefore not modeled in our simulator as a result.}.

    As mentioned, dynamic traversal workloads require extensive use of fine-grained synchronization.
    Moreover, these algorithms require the values returned by these racy accesses in order to determine the traversal's control flow.
    Since the values for these racy accesses are being read, diligent reasoning by the programmer should be used to determine whether these racy accesses can be relaxed using DRFrlx according to the specification described in~\cite{SinclairAlsop2017}.
    Additionally, a consequence of reading the values returned by racy accesses for control flow is that a thread must wait for the values to proceed computation, limiting the impact of atomic relaxation on performance.
    Therefore, specialize for a consistency model that offers ease of programmability~(DRF1) without considerable loss in performance.
    
    \subsection{Partial Design Space Optimization}
    \label{subsec:model-partial}
    The model described above is not sufficient to reason about the optimal design given design space restrictions.
    Namely, if DRFrlx is not available on the target GPU, the design choice between push and pull becomes more complicated.
    If DRFrlx is not available, there is naturally no longer a choice in the consistency dimension.
    Furthermore, we always recommend GPU coherence if pull is selected, and the coherence decision is made independently of the consistency decision.
    It follows that the only decision to be reconsidered is when to push or pull.
    
    Similar to the full model, if AC elides more work at source we opt for a push implementation.
    Because the amount of loads that can be hoisted via AI is generally less than the amount of work elided by AC, this is broken off into a second-order decision to be considered only if AC does not prefer source.
    If AI prefers source, then we proceed to the same secondary push decision as the full model (medium/low locality, etc.\ ).
    If AI does not prefer source, then we consider stricter requirements to recommend push; medium volume is no longer sufficient for push, it must be high.
    This is because in this case, the relatively small affects of medium volume on cache thrashing for pull generally do not outweigh the costs of atomics (as they can not be relaxed in this model).

\section{Methodology}
\label{sec:methodology}

\begin{table*}[!t]
    \centering
    \captionsetup{font=footnotesize}
    \caption{Input graph statistics and their taxonomy classifications.  H indicates a high value, M for medium, and L for low.
    \textcolor{blue}{Blue} and \textcolor{red}{red} classifications indicated favorable and negative characteristics respectively.}
    \scriptsize
    \begin{tabular}{l | r r r r r r r r r r}
        \toprule
        \textbf{Graph} & \textbf{Vertices} & \textbf{Edges} & \textbf{Max Deg} & \textbf{Avg Deg} & \textbf{Std Dev Def} & \textbf{Volume (KB)} & $\mathbf{AN_L}$ & $\mathbf{AN_R}$ & \textbf{Reuse} & \textbf{Imbalance} \\
        \midrule
            AMZ & 410236 & 6713648 & 2770  & 16.265 & 16.298  & \textcolor{red}{1855.178 (H)}   & 2.616 & 13.749 & \textcolor{orange}{0.160 (M)} & \textcolor{blue}{0.000 (L)} \\
            DCT & 52652  & 178076  & 38    & 3.382  & 4.475   & \textcolor{orange}{60.078 (M)}  & 1.215 & 2.167  & \textcolor{orange}{0.359 (M)} & \textcolor{orange}{0.083 (M)} \\
            EML & 265214 & 837912  & 7636  & 3.159  & 42.490  & \textcolor{red}{287.272 (H)}    & 0.167 & 2.992  & \textcolor{red}{0.053 (L)}    & \textcolor{red}{1.000 (H)} \\
            OLS & 88263  & 683186  & 10    & 7.740  & 2.411   & \textcolor{orange}{200.898 (M)} & 3.446 & 4.295  & \textcolor{blue}{0.445 (H)}   & \textcolor{blue}{0.000 (L)} \\
            RAJ & 20640  & 163178  & 3469  & 7.906  & 32.954  & \textcolor{blue}{47.869 (L)}    & 4.697 & 3.209  & \textcolor{blue}{0.594 (H)}   & \textcolor{red}{0.617 (H)} \\
            WNG & 61032  & 243088  & 4     & 3.919  & 0.278   & \textcolor{orange}{79.458 (M)}  & 0.020 & 3.899  & \textcolor{orange}{0.594 (L)} & \textcolor{blue}{0.000 (L)} \\
            \bottomrule
    \end{tabular}
    \label{tbl:graphs}%
\end{table*}
\begin{table}[!ht]
  \centering
  \captionsetup{font=footnotesize}
  \caption{Algorithmic properties for each application.
           The `\okay' markers for the CC application indicate where properties are dependent on the dynamic behavior of the workload, and thus not used for specialization.}%
  \scriptsize
  \begin{tabular}{ c | c c c }
    \toprule
      \textbf{App}        & \textbf{Traversal}  & \textbf{Control}  & \textbf{Information}    \\
    \midrule                
    PR                    & Static              & Symmetric         & Source                  \\
    SSSP                  & Static              & Source            & Source                  \\
    MIS                   & Static              & Symmetric         & Symmetric               \\
    CLR                   & Static              & Symmetric         & Target                  \\
    BC                    & Static              & Source            & Symmetric               \\
    CC                    & Dynamic             & \okay             & \okay                   \\
    \bottomrule
  \end{tabular}%
  \normalsize
  \label{tbl:apps}%
\end{table}%

\begin{table}[!t]
  \centering
  \captionsetup{font=footnotesize}
  \caption{Simulated heterogeneous system parameters.}%
  \scriptsize
  \begin{tabular}{ c c }
    \toprule
    \multicolumn{2}{c}{\textbf{CPU Parameters}} \\
    \midrule
    Frequency                       & 2 GHz \\ 
    Cores                           & 1 \\
    \toprule
    \multicolumn{2}{c}{\textbf{GPU Parameters}} \\
    \midrule
    Frequency                       & 700 MHz \\ 
    CUs                             & 15 \\
    \toprule
    \multicolumn{2}{c}{\textbf{Memory Hierarchy Parameters}} \\
    \midrule
    L1 Size (8 banks, 8-way assoc.) & 32 KB \\ 
    L2 Size (16 banks, NUCA)        & 4 MB \\ 
    Store Buffer Size               & 128 entries \\ 
    L1 MSHRs                        & 128 entries \\ 
    L1 hit latency                  & 1 cycle \\ 
    Remote L1 hit latency           & 35--83 cycles \\ 
    L2 hit latency                  & 29--61 cycles\\ 
    Memory latency                  & 197--261 cycles \\
    \bottomrule
  \end{tabular}%
  \normalsize
  \label{tbl:sim}%
\end{table}%

  \subsection{Graph Inputs}
  \label{subsec:methodology-graph-inputs}

  We select six representative input graphs from the SuiteSparse matrix collection~\cite{DavisHu2011} with a variety of connectivity features.
  Where applicable, each graph has been slightly modified to remove self-edges, and has been converted to a directed, symmetric graph to support push and pull kernels using the same input.
  Table~\ref{tbl:graphs} lists the name, number of vertices and edges, and other basic properties of these graphs.

  For each graph, we compute the volume, reuse, and imbalance metrics described in Sections~\ref{subsubsec:taxonomy-volume}, \ref{subsubsec:taxonomy-reuse} and \ref{subsubsec:taxonomy-imbalance}, as shown in Table~\ref{tbl:graphs}.
  For each graph, the graph structure profile is subsequently used in our workload-driven methodology to predict the optimal design to specialize to for a given workload.


The various thresholds we choose are: (a) low volume - less than 1.5 times the L1 data cache size, high volume - greater than the L2 cache size divided by the number of GPU SMs, (b) low reuse - less than 0.15, high reuse - larger than 0.40
~(for the $Reuse$ formula)
, (c) low imbalance - less than 0.05, high imbalance - larger than 0.25
~(for the $Imbalance$ metric)
and (d) k-means clustering threshold~(Section~\ref{subsubsec:taxonomy-imbalance}) for the max degree centroid differential is 10.
  These thresholds were empirically determined based on the cache performance of our experiments on a larger set of graphs than evaluated in Section~\ref{sec:results}.
  This classifier is rather simple, but the focus of this work is on the interaction of these graph attributes, algorithmic properties, and the consistency and coherence of a system; alternative classifiers would also be sufficient.

  \subsection{Applications}
  \label{subsec:methodology-applications}

  We evaluate six applications (for each of the graph inputs from the previous section). Five of them are adapted from the Pannotia benchmark suite~\cite{CheBeckmann2013} and exhibit a wide range of algorithmic control and information.
  These are PageRank~(PR), Single-Source Shortest Path~(SSSP), Maximal Independent Set~(MIS), Graph Coloring~(CLR), and Betweenness Centrality~(BC).
  We have developed new implementations where either a push or pull version of the application did not exist.
  Furthermore, we have modified the kernels by unrolling inner loops and inlining hand-written assembly code to show the benefits of relaxing the ordering of atomic operations within a thread~\cite{SinclairAlsop2017}.
  A sixth application, Connected Components~(CC), is adapted from~\cite{JaiganeshBurtscher2018} to represent the design space for workloads with dynamic algorithmic traversal.
  We find that these applications are similar to those found in contemporary works on graph processing and irregular memory accesses on GPUs~\cite{Li2019,Xu2019}.
  The algorithmic properties described in Section~\ref{sec:taxonomy}, used to characterize each application, are summarized in Table~\ref{tbl:apps}; their values having been determined by manual inspection of application code.

  \subsection{Simulation Infrastructure}
  \label{subsec:methodology-hardware}

  We simulate a tightly-integrated, coherent CPU-GPU system using the same simulator as prior work, obtained from the authors~\cite{SinclairAlsop2017}.
  This simulator uses GEMS~\cite{MartinSorin2005} to model a coherent memory system, Garnet~\cite{AgarwalKrishna2009} to model a 4x4 mesh network, Simics~\cite{MagnussonChristensson2002} to model the CPU, and GPGPU-Sim~\cite{BakhodaYuan2009} to model the GPU.
  This work must be simulated (as opposed to studied on a real machine) since there are no means to change the coherence  and memory consistency of conventional CPU-GPU systems to capture the design space range covered here.
  Each core has a private L1 cache, and all cores shared a banked L2 cache.
  We summarize the key hardware parameters in Table~\ref{tbl:sim}.
  
  We measure the execution time of each graph workload using a stall classification methodology similar to that described by Alsop et al.~\cite{AlsopSinclair2016}.
  \textit{Busy} represents time spent doing useful work~(cycles where 1 or more instructions were able to issue). 
  \textit{Comp} represents time spent waiting for an occupied computation unit or the result of a computation.
  \textit{Data} represents time spent waiting for an occupied LD/ST unit, or waiting for the result of a non-atomic memory operation.
  \textit{Sync} represents time spent waiting for the result of an atomic memory operation, or waiting at a thread barrier.
  \textit{Idle} represents idle time spent by a GPU core as it waits for other cores to complete a kernel.
  
  \subsection{Configurations}
  \label{subsec:methodology-configuration}
  We evaluate each workload across the full suite of possible optimizations.
  In each configuration, from left to right: $T$ and $S$ denote Target and Source updates, respectively, in static traversal, and $D$ denotes Dynamic traversal; $G$ and $D$ denote GPU and DeNovo coherence, respectively; $0$, $1$, and $R$ denote DRF0, DRF1, and DRFrlx, respectively.
\section{Evaluation}
\label{sec:results}


\begin{table}[!t]
  \centering
  \captionsetup{font=footnotesize}
  \caption{Each entry in the table is the design predicted as best by our model for a given workload (the column gives the algorithm and the row gives the input).
The entries in \close{gray} are mispredicted but perform within 3.5\% of the actual best design.
  }%
  \scriptsize
  \begin{tabular}{ c | c | c | c | c | c | c }
      \toprule
                   & \textbf{PR}                 & \textbf{SSSP}             & \textbf{MIS}              & \textbf{CLR}          & \textbf{BC}               & \textbf{CC}               \\
      \midrule
      \textbf{AMZ} & SGR                         & SGR                       & SGR                       & SGR                   & SGR                       & \close{DD1}                   \\
      \textbf{DCT} & SGR                         & SGR                       & SGR                       & SGR                   & SGR                       & \close{DD1}                   \\
      \textbf{EML} & SGR                         & \close{SGR}               & SGR                       & SGR                   & SGR                       & \close{DD1}           \\
      \textbf{OLS} & SDR                         & SDR                       & TG0                       & TG0                   & \close{SDR}               & \close{DD1}                   \\
      \textbf{RAJ} & SDR                         & SDR                       & SDR                       & SDR                   & SDR                       & \close{DD1}                   \\
      \textbf{WNG} & SGR                         & \close{SGR}               & SGR                       & SGR                   & SGR                       & DD1                   \\
    \bottomrule
   \end{tabular}%
   \normalsize
  \label{tbl:designs}%
\vspace{-.20in}
\end{table}%

\begin{figure*}[!ht]
    \setlength{\columnsep}{0.0pt}
    \setlength{\multicolsep}{0.0pt}
    \centering
    \begin{multicols}{2}
        \includegraphics[keepaspectratio,width=\linewidth]{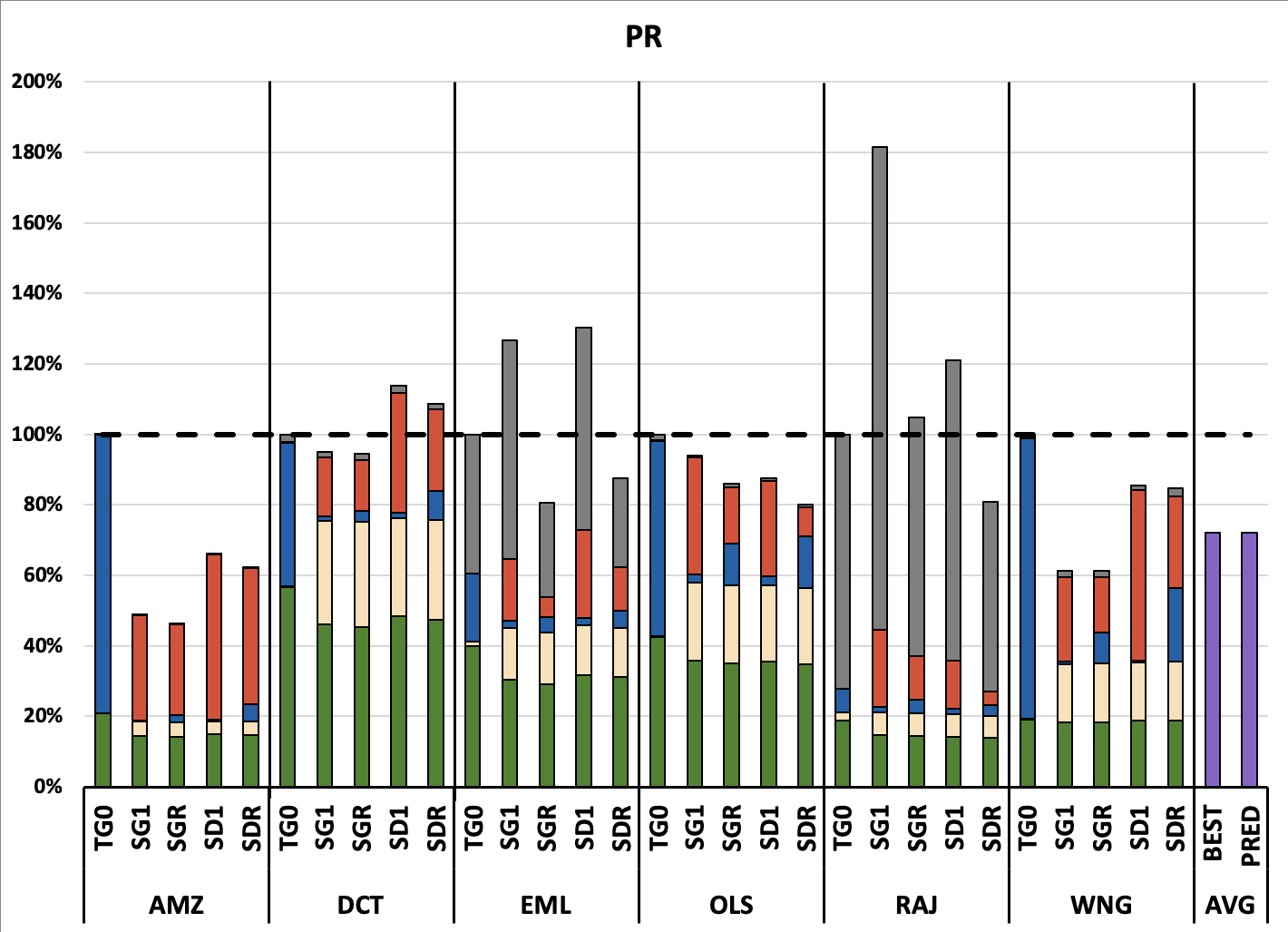}\par
        \includegraphics[keepaspectratio,width=\linewidth]{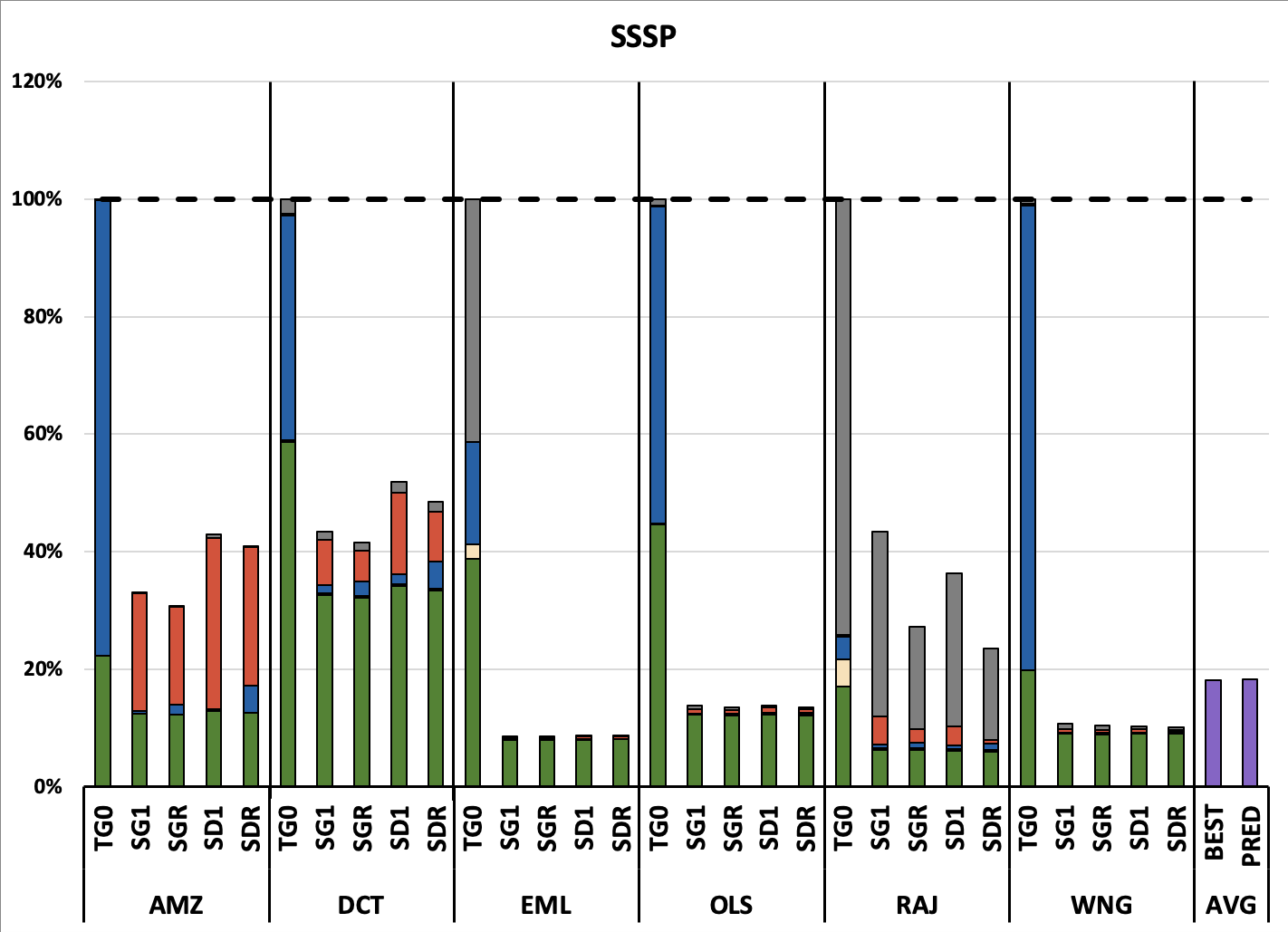}\par
    \end{multicols}%
    \begin{multicols}{2}
        \includegraphics[keepaspectratio,width=\linewidth]{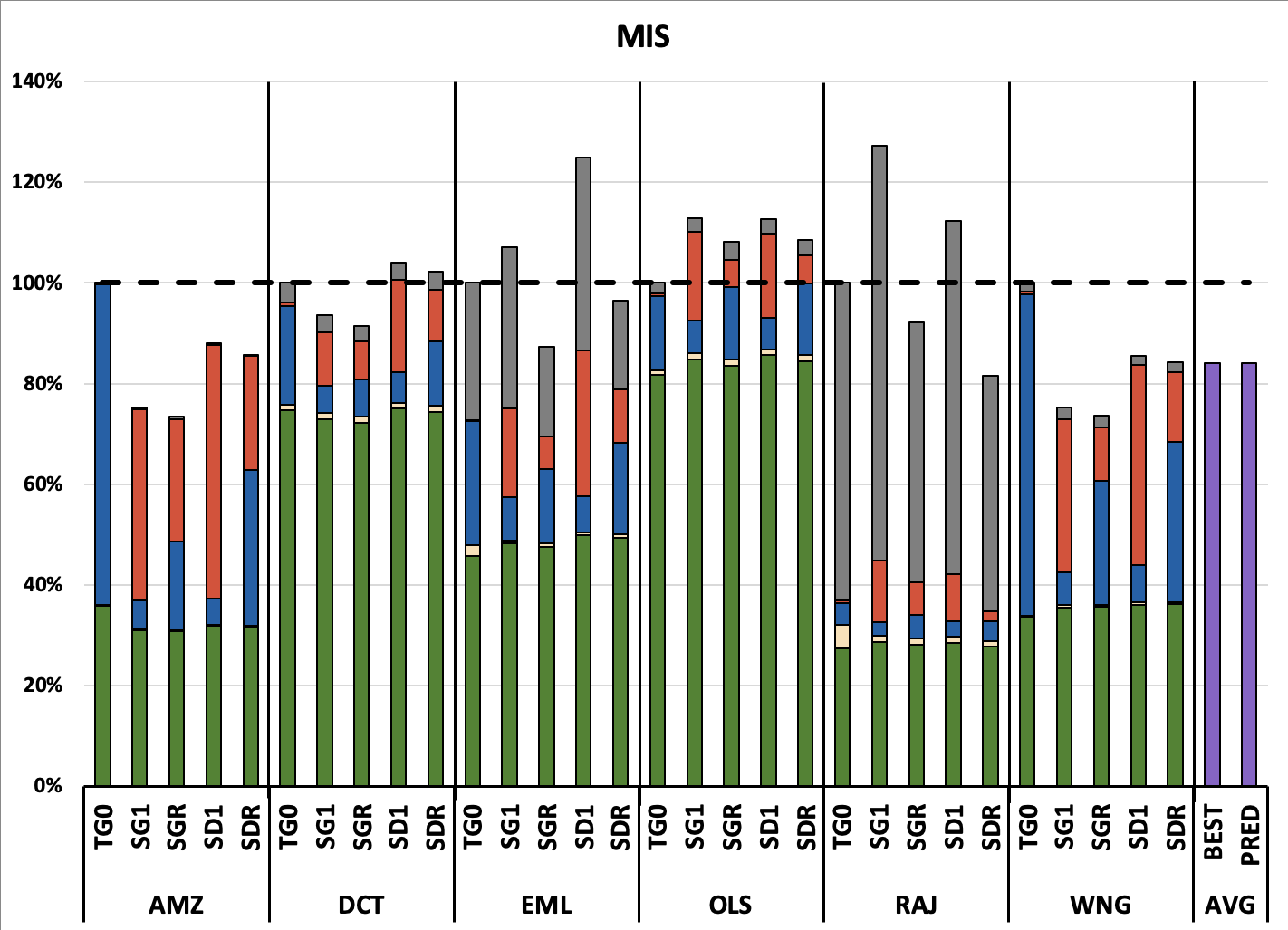}\par
        \includegraphics[keepaspectratio,width=\linewidth]{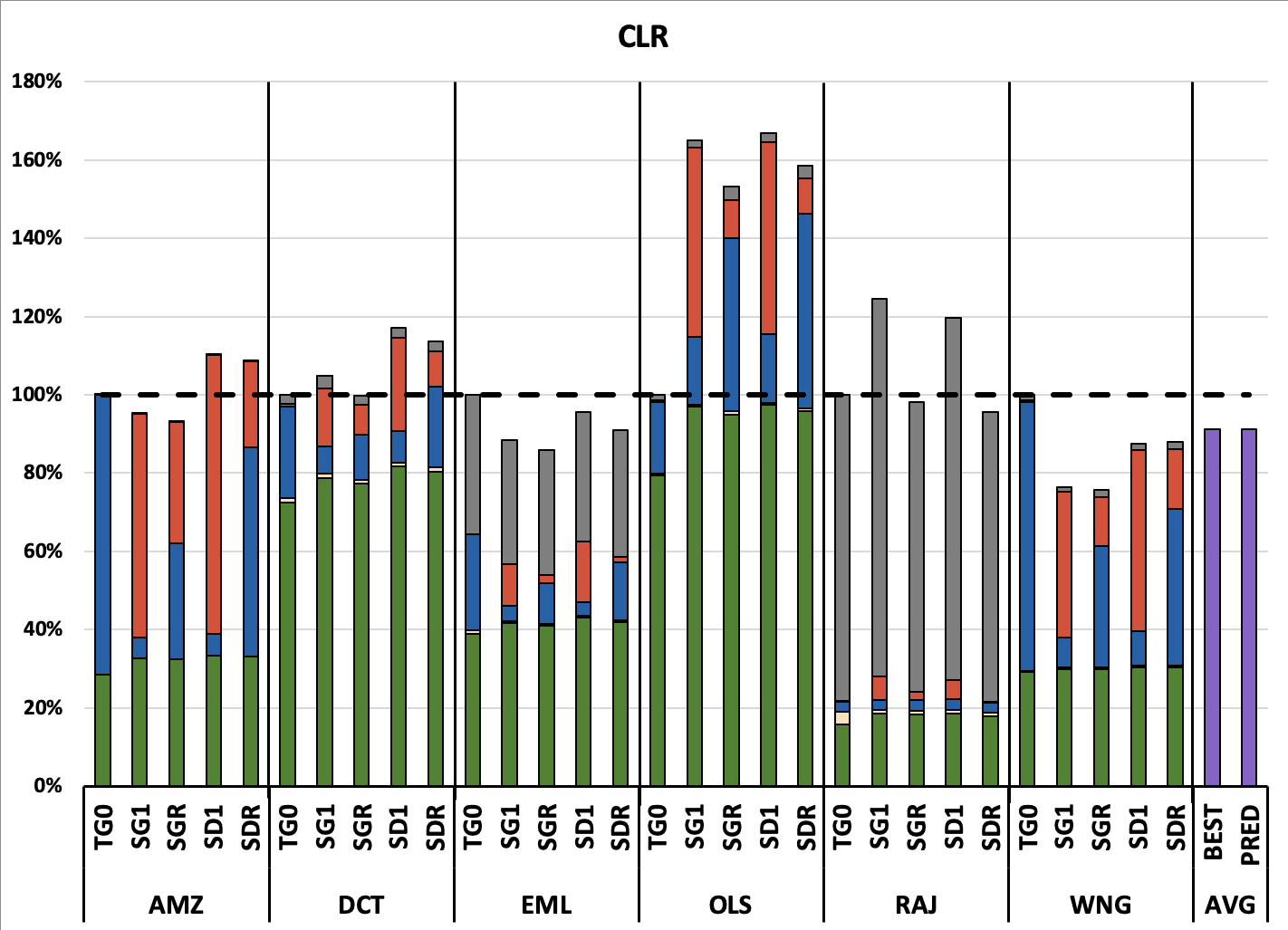}
    \end{multicols}%
    \begin{multicols}{2}
        \includegraphics[keepaspectratio,width=\linewidth]{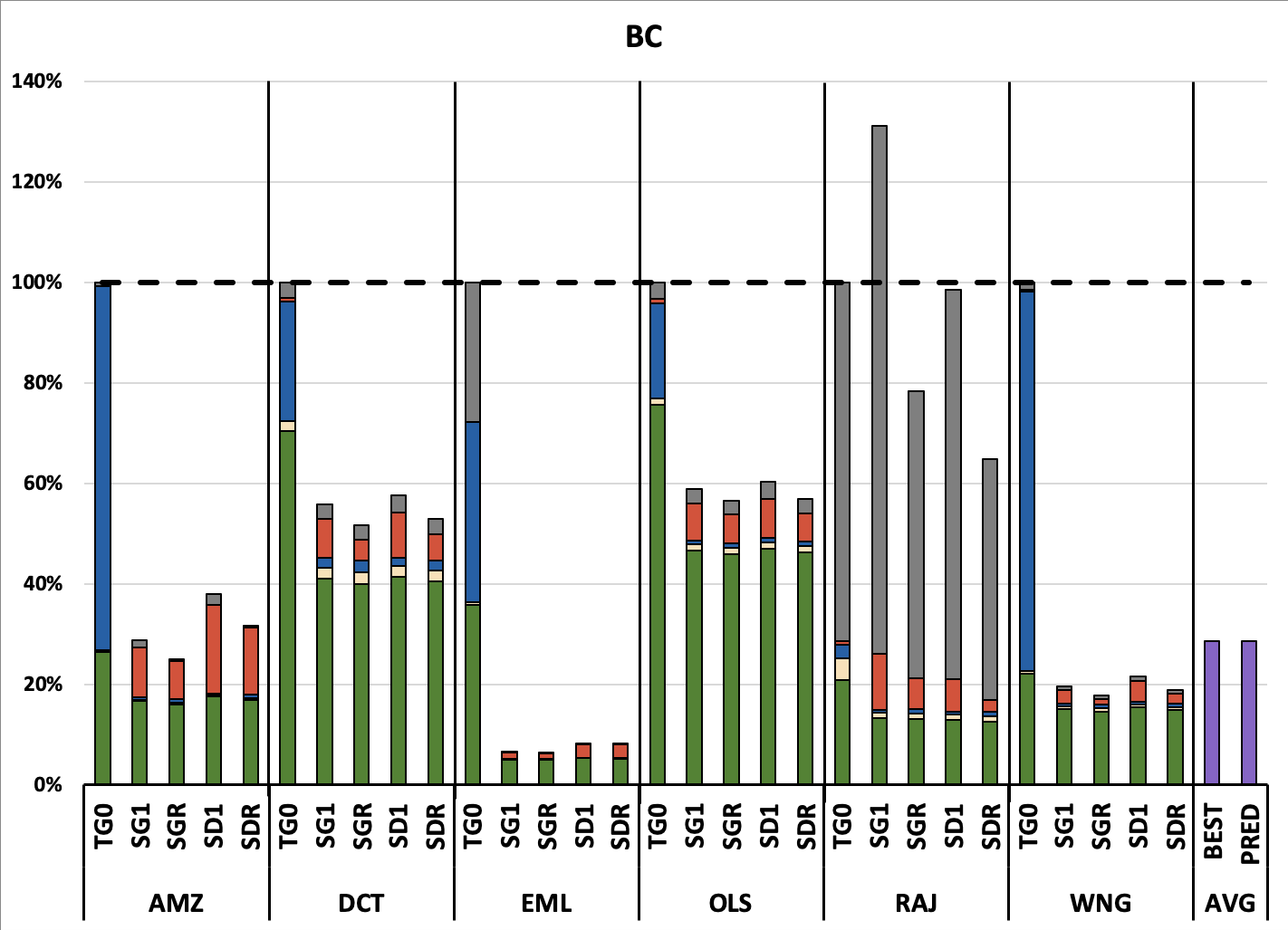}\par
        \includegraphics[keepaspectratio,width=\linewidth]{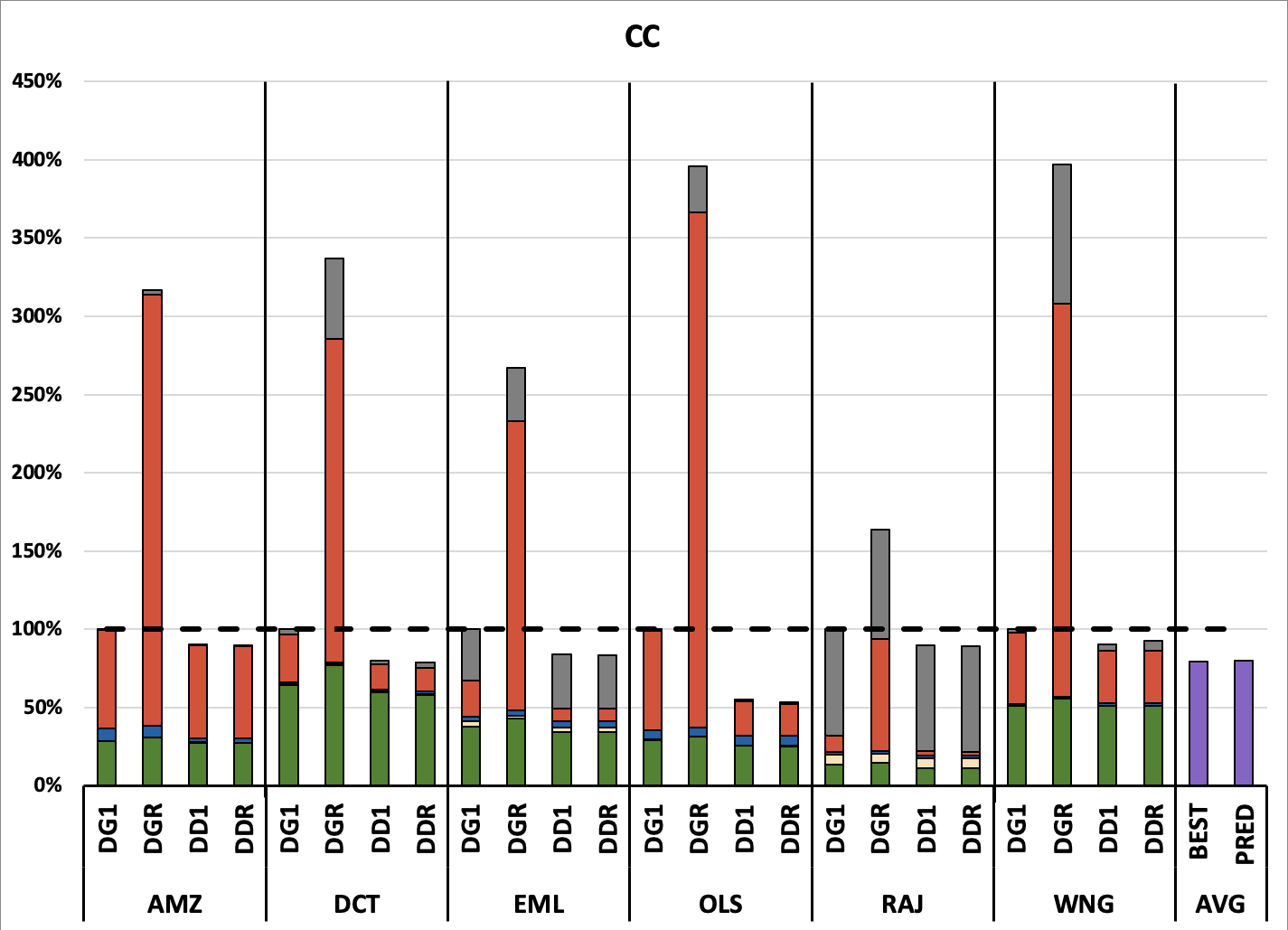}\par
    \end{multicols}%
    \includegraphics[keepaspectratio,width=0.6\textwidth]{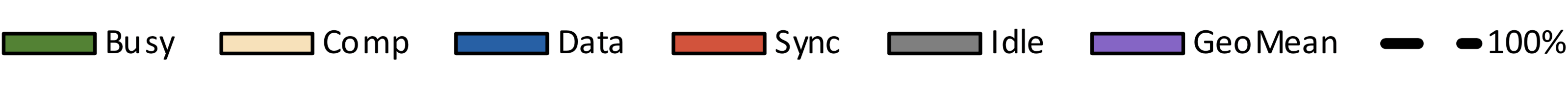}%
    \captionsetup{font=footnotesize}
    \caption{GPU execution time breakdown for six different graph applications, each with six different input graphs.
             All bars of a particular input are normalized to the leftmost configuration --- DG1 for CC, TG0 for all other applications.
             Each application also has two bars shown for the geometric mean performance of designs across inputs: BEST for empirically found optimal designs, and PRED for designs predicted by our methodology. In each configuration, from left to right: $T$ and $S$ denote Target and Source updates, respectively, in static traversal, and $D$ denotes Dynamic traversal; $G$ and $D$ denote GPU and DeNovo coherence, respectively; $0$, $1$, and $R$ denote DRF0, DRF1, and DRFrlx, respectively.}%
    \label{fig:results}%
\vspace{-.15in}
\end{figure*}%

\begin{figure}
\centering
\includegraphics[keepaspectratio,width=\columnwidth]{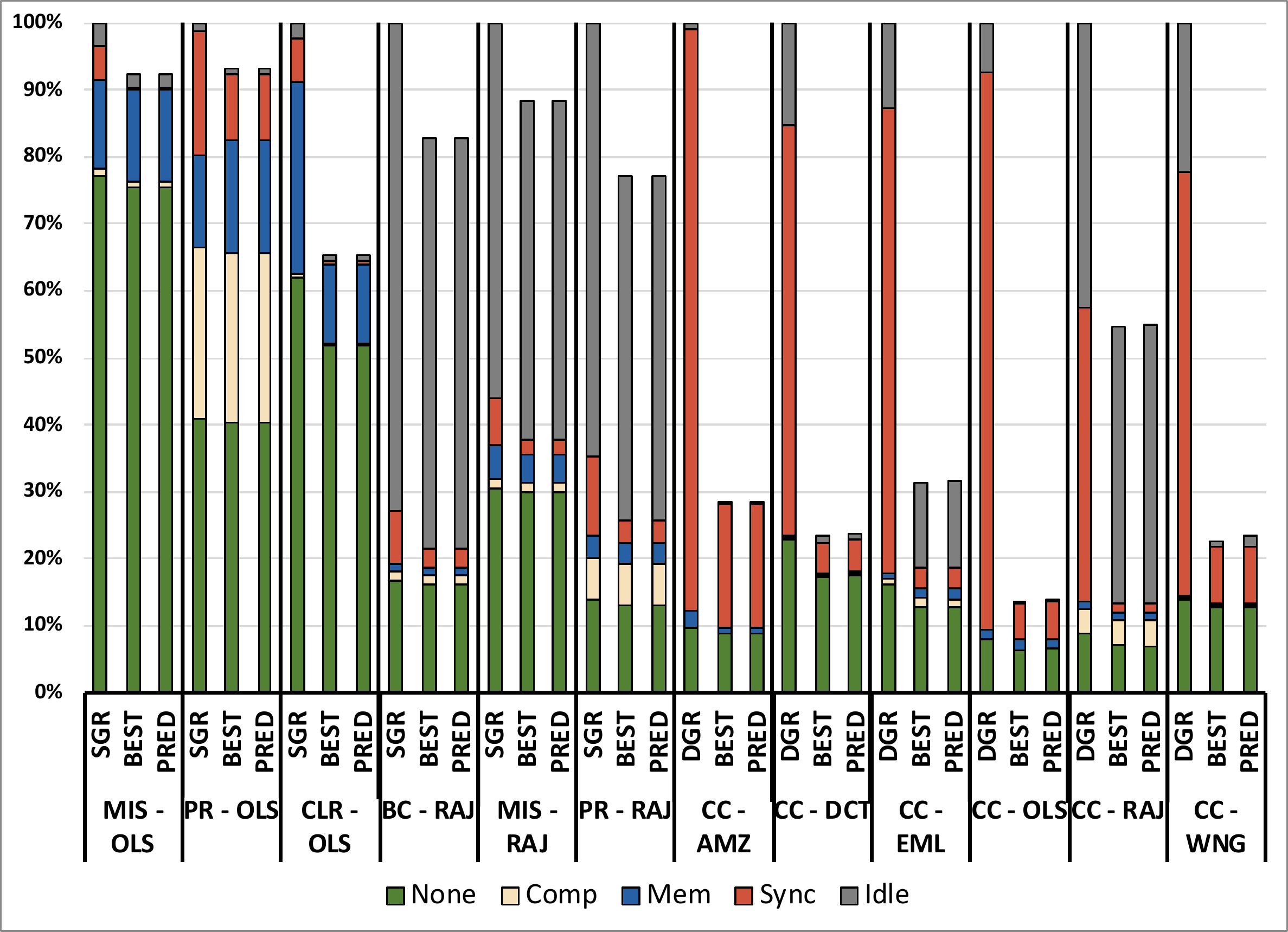}
\captionsetup{font=footnotesize}
\caption{Normalized execution time comparison of SGR (DGR where applicable) to the empirical best and predicted designs}
\label{fig:cgr}
\vspace{-.25in}
\end{figure}


Figure~\ref{fig:results} shows the results for the 6 graph applications with 6 inputs per application. For all but the CC application, for each input, we show execution times for 5 design configurations, normalized to the Pull configuration, TG0 (for target, GPU, and DRF0). Since Pull does not use fine-grained atomics, it does not show performance variation with the consistency models or coherence protocols. We therefore show only one Pull configuration (the simplest one, TG0). Further, DRF0 performs poorly (relative to DRF1/DRFrlx) for all Push configurations since these configurations have very significant use of atomics and DRF0 incurs large atomics overhead. For brevity, we therefore show data only for DRF1 and DRFrlx for Push, resulting in a total of 5 configurations shown for each workload. For CC, we do not have a choice of pull or push (the application uses both), reducing the total configurations shown to four per graph input for CC. 

Our results have three key takeaways.

{\bf Need for flexibility -- one size does not fit all:}
A primary take away from our results is that there is no single hardware/software configuration that works best for all workloads.
While SGR works best for many of these workloads, there are 12 cases where it does not, summarized in Figure~\ref{fig:cgr}.
In these cases we see that relative to SGR, the best configuration reduces execution time ranging from 7\% to 87\%, averaging a 44\% reduction.
Also note that while we say BEST for clarity, the best configuration varies workload to workload.

These results motivate the need for flexible hardware mechanisms that support flexible coherence and consistency models. Recent work such as Spandex~\cite{AlsopSinclair2018} that shows how to efficiently integrate different coherence protocols in heterogeneous systems provides a path to such flexibility while preserving the simplicity and efficiency desired by such systems.

{\bf The model is accurate:}
Another key takeaway is that our predictive model (as described in Section~\ref{subsec:model-full}) accurately predicts the optimal HW/SW configuration in 28 out of 36 workloads.
All 8 mispredictions perform within $3.5$\% of the empirical best, and average $1.3$\% difference.
Five of these workloads ((AMZ,CC), (DCT,CC), (EML,CC), (OLS,CC), (RAJ,CC)) had predictions that were expected to be sub-optimal and chosen to relieve burden on the programmer in exchange for a minimal reduction of performance (average $1.3$\%) as described in our model.
The other three workloads ((EML,SSSP), (OLS,BC), (WNG,SSSP)) mispredict as the result of known limitations of our metrics.
We will consider (EML,SSSP) as an example, but similar reasoning can be applied to the other workloads as well.

EML is a graph that is categorized as high volume, low reuse, and high imbalance.  SSSP is an application that elides more work at source and hoists more information at source.  Following our decision flow, the optimal configuration for this should be SGR; however, empirically, the optimal is SG1 (albeit with a very small difference from SGR). The reason for the (small) difference is that we are limited by our static metrics.
The static volume and imbalance calculations do not consider the run time affects the application will have on these attributes.
The first kernel of SSSP is a breadth first search.  EML's power-law distribution of edges means the working set size (approximated by our volume calculation) is actually quite small, despite the large size of the graph.  Similarly, if one of the high degree nodes is not on the frontier, the imbalance for this workload will be quite low.  A decision flow similar to our model that used runtime information could consider this and choose the correct configuration.

{\bf Inter-dependent design dimensions:}
Finally, the results show that the design dimensions considered in this work are inter-dependent. 
For example, consider MIS and RAJ.  If the system does not support DRFrlx, then the best configuration is Pull (TG0). If the system supports DRFrlx, then the best configuration is Push with SDR. This shows that the design decision of Pull vs.\ Push cannot be made independent of the knowledge of the consistency model supported by the system.
Overall, there are seven workloads where without efficient support for DRFrlx, we would recommend a pull implementation over a push one.
Applying the extension of the model to the partial design space (Section~\ref{subsec:model-partial}), we find that we are able to accurately predict these decisions for four of the seven cases. For the other three cases, we need finer grain distinctions in the volume and AI input parameters -- we omit these extensions here for lack of space.  

\section{Related Work}
\label{sec:related}

  \subsection{Graph Analytics Frameworks}
  \label{subsec:related-frameworks}

  There exist many programmable frameworks for implementing graph analytics workloads~\cite{NguyenLenharth2013,WangPan2017,ShunBlelloch2013,BurtscherNasre2012,GrossmanLitz2018,SundaramSatish2015,RoyMihailovic2013,ZhangYang2018,HongChafi2012,ZhangChen2015,KyrolaBlelloch2012,KhorasaniVora2014,SabetQiu2018}.
  All of these frameworks identify the various and diverse demands of graph analytics applications on the processing sub-systems, including data structure representations, scheduling and compute-to-data mapping, structural transformations of data, and data access type annotations to improve locality and performance and to reduce communication overheads.
  However, there are several limitations of these works as they relate to the design space explored in our work.
  Most of the works on graph processing frameworks focus on CMP systems using commodity hardware, some focusing on distributed systems, while others on shared memory.
  Our work explores the implications of future heterogeneous systems with unified, coherent shared memory and flexible coherence and consistency.
  For works that have explored GPU implementations, none have considered GPUs with fully coherent memory systems on a tightly-coupled CPU-GPU system with unified, shared memory.
  Moreover, none of these works have considered nor characterized the design space of coherence and consistency for GPUs, especially in the context of data-dependent and algorithmic-dependent behavior of graph workloads.


  \subsection{Synchronization}
  \label{subsec:related-sync}

  As the field of graph analytics develops, more works have explored the implications of racy (push-style) updates and their relationship with synchronization.
  Many works only consider either push- or pull-style implementations or do not have comprehensive comparison for several algorithms~\cite{SabetQiu2018,GrossmanLitz2018}.
  Some explore both pushing and pulling, and even the idea of choosing the best implementation at runtime as the state of the graph processing progresses~\cite{BestaPodstawski2017,WangPan2017,ShunBlelloch2013,ZhangYang2018}, but do not consider trade-offs of racy and non-racy updates in the presence of coherence and consistency specialization.
  Some works identify that existing systems perform poorly for graph analytics workloads due to the fine-grained synchronization overheads using atomics~\cite{NasreBurtscher2013,ElteirLin2011,XiaoFeng2010,GrossmanLitz2018,KaleemVenkat2016}, but do not consider architectures with efficient coherence and consistency support for atomics.
  Previous works that explore the design space of coherence and consistency for heterogeneous CMPs~\cite{OrrChe2015,SinclairAlsop2015,AlsopOrr2016,AlsopSinclair2018} do not 
   consider how these trade-offs change with respect to pushing and pulling.
  Finally, works on programmable architectures for graph analytics~\cite{HamWu2016,PrabhakarZhang2017} have more rigid memory systems with little to no fine-grained synchronization support, and do not implement coherence because the accelerators do not operate in a unified, global address space.







  \subsection{Input Sensitivity and Concurrency}
  \label{subsec:related-input-sensitivity}

  Several works have explored the relationship of data movement, input sensitivity, and concurrency on the performance of both CPU and GPU systems~\cite{BurtscherNasre2012,BestaPodstawski2017,AhmadKhan2016,HongOguntebi2011,LiBecchi2013,HestnessKeckler2015,XuJeon2014,EyermanHeirman2018,Ahmad2019,Sorensen2019}. 
  However, these works do not consider either the role of the compute-to-data mapping or coherence and consistency on data- and algorithmic-dependent performance scaling.
  \cite{EyermanHeirman2018} does explore the interplay of coherence and data-dependence, but does not explore consistency models.




  


\section{Conclusion}
\label{sec:conclusion}

  Our taxonimization and evaluation of graph analytics in relation to update propagation, consistency, and coherence has shown that the design space for these workloads leads to non-trivial and inter-dependent design decisions that are input- and algorithm-dependent.
  The optimal configurations and the performance of each configuration vary widely by workload.
  Based on our insights, we have proposed a specialization model that accurately predicts the optimal software-hardware configuration in 28 out of 36 workloads, and predicts a configuration that is within 3.5\% in the remaining 8 workloads.
 We also show that we can extend the model to apply our taxonomy to a restricted design space.
  Using our model, software developers can make hardware-aware design choices and hardware designers can leverage the flexibility and reconfigurability of upcoming memory systems~(e.g., \cite{AlsopSinclair2018}) to specialize for diverse irregular graph workloads on configurable processor architectures.
  Looking forward, we aim to target our analysis to implement runtime methods that leverage flexible memory systems to achieve optimal performance, and to extend our taxonomy to other classes of algorithms and graph datasets.





\bibliographystyle{IEEEtran}
\bibliography{bibliography/hetero.bib}

\begin{thebibliography}{10}
\providecommand{\url}[1]{#1}
\csname url@samestyle\endcsname
\providecommand{\newblock}{\relax}
\providecommand{\bibinfo}[2]{#2}
\providecommand{\BIBentrySTDinterwordspacing}{\spaceskip=0pt\relax}
\providecommand{\BIBentryALTinterwordstretchfactor}{4}
\providecommand{\BIBentryALTinterwordspacing}{\spaceskip=\fontdimen2\font plus
\BIBentryALTinterwordstretchfactor\fontdimen3\font minus
  \fontdimen4\font\relax}
\providecommand{\BIBforeignlanguage}[2]{{%
\expandafter\ifx\csname l@#1\endcsname\relax
\typeout{** WARNING: IEEEtran.bst: No hyphenation pattern has been}%
\typeout{** loaded for the language `#1'. Using the pattern for}%
\typeout{** the default language instead.}%
\else
\language=\csname l@#1\endcsname
\fi
#2}}
\providecommand{\BIBdecl}{\relax}
\BIBdecl

\bibitem{Eeckhout2017}
L.~Eeckhout, ``Is moore's law slowing down? what's next?'' \emph{IEEE Micro},
  vol.~37, no.~4, pp. 4--5, 2017.

\bibitem{NguyenLenharth2013}
D.~Nguyen, A.~Lenharth, and K.~Pingali, ``A lightweight infrastructure for
  graph analytics,'' in \emph{Proceedings of the Twenty-Fourth ACM Symposium on
  Operating Systems Principles}, ser. SOSP '13.\hskip 1em plus 0.5em minus
  0.4em\relax New York, NY, USA: ACM, 2013, pp. 456--471.

\bibitem{WangPan2017}
\BIBentryALTinterwordspacing
Y.~Wang, Y.~Pan, A.~A. Davidson, Y.~Wu, C.~Yang, L.~Wang, M.~Osama, C.~Yuan,
  W.~Liu, A.~T. Riffel, and J.~D. Owens, ``Gunrock: {GPU} graph analytics,''
  \emph{CoRR}, vol. abs/1701.01170, 2017. [Online]. Available:
  \url{http://arxiv.org/abs/1701.01170}
\BIBentrySTDinterwordspacing

\bibitem{ShunBlelloch2013}
J.~Shun and G.~E. Blelloch, ``Ligra: A lightweight graph processing framework
  for shared memory,'' in \emph{Proceedings of the 18th ACM SIGPLAN Symposium
  on Principles and Practice of Parallel Programming}, ser. PPoPP '13.\hskip
  1em plus 0.5em minus 0.4em\relax New York, NY, USA: ACM, 2013, pp. 135--146.

\bibitem{BurtscherNasre2012}
M.~Burtscher, R.~Nasre, and K.~Pingali, ``{A quantitative study of irregular
  programs on GPUs},'' in \emph{{2012 IEEE International Symposium on Workload
  Characterization}}, ser. {IISWC}, Nov 2012, pp. 141--151.

\bibitem{GrossmanLitz2018}
S.~Grossman, H.~Litz, and C.~Kozyrakis, ``Making pull-based graph processing
  performant,'' in \emph{Proceedings of the 23rd ACM SIGPLAN Symposium on
  Principles and Practice of Parallel Programming}, ser. PPoPP '18.\hskip 1em
  plus 0.5em minus 0.4em\relax New York, NY, USA: ACM, 2018, pp. 246--260.

\bibitem{SundaramSatish2015}
N.~Sundaram, N.~Satish, M.~M.~A. Patwary, S.~R. Dulloor, M.~J. Anderson, S.~G.
  Vadlamudi, D.~Das, and P.~Dubey, ``Graphmat: High performance graph analytics
  made productive,'' \emph{Proc. VLDB Endow.}, vol.~8, no.~11, pp. 1214--1225,
  Jul. 2015.

\bibitem{RoyMihailovic2013}
A.~Roy, I.~Mihailovic, and W.~Zwaenepoel, ``X-stream: Edge-centric graph
  processing using streaming partitions,'' in \emph{Proceedings of the
  Twenty-Fourth ACM Symposium on Operating Systems Principles}, ser. SOSP
  '13.\hskip 1em plus 0.5em minus 0.4em\relax New York, NY, USA: ACM, 2013, pp.
  472--488.

\bibitem{ZhangYang2018}
\BIBentryALTinterwordspacing
Y.~Zhang, M.~Yang, R.~Baghdadi, S.~Kamil, J.~Shun, and S.~Amarasinghe,
  ``Graphit: A high-performance graph dsl,'' \emph{Proc. ACM Program. Lang.},
  vol.~2, no. OOPSLA, pp. 121:1--121:30, Oct. 2018. [Online]. Available:
  \url{http://doi.acm.org/10.1145/3276491}
\BIBentrySTDinterwordspacing

\bibitem{HongChafi2012}
S.~Hong, H.~Chafi, E.~Sedlar, and K.~Olukotun, ``Green-marl: A dsl for easy and
  efficient graph analysis,'' in \emph{Proceedings of the Seventeenth
  International Conference on Architectural Support for Programming Languages
  and Operating Systems}, ser. ASPLOS XVII.\hskip 1em plus 0.5em minus
  0.4em\relax New York, NY, USA: ACM, 2012, pp. 349--362.

\bibitem{ZhangChen2015}
K.~Zhang, R.~Chen, and H.~Chen, ``Numa-aware graph-structured analytics,'' in
  \emph{Proceedings of the 20th ACM SIGPLAN Symposium on Principles and
  Practice of Parallel Programming}, ser. PPoPP 2015.\hskip 1em plus 0.5em
  minus 0.4em\relax New York, NY, USA: ACM, 2015, pp. 183--193.

\bibitem{KyrolaBlelloch2012}
A.~Kyrola, G.~Blelloch, and C.~Guestrin, ``Graphchi: Large-scale graph
  computation on just a pc,'' in \emph{Proceedings of the 10th USENIX
  Conference on Operating Systems Design and Implementation}, ser.
  OSDI'12.\hskip 1em plus 0.5em minus 0.4em\relax Berkeley, CA, USA: USENIX
  Association, 2012, pp. 31--46.

\bibitem{KhorasaniVora2014}
F.~Khorasani, K.~Vora, R.~Gupta, and L.~N. Bhuyan, ``{CuSha: Vertex-centric
  Graph Processing on GPUs},'' in \emph{{Proceedings of the 23rd International
  Symposium on High-performance Parallel and Distributed Computing}}, ser.
  {HPDC}, 2014, pp. 239--252.

\bibitem{SabetQiu2018}
A.~H. Nodehi~Sabet, J.~Qiu, and Z.~Zhao, ``Tigr: Transforming irregular graphs
  for gpu-friendly graph processing,'' in \emph{Proceedings of the Twenty-Third
  International Conference on Architectural Support for Programming Languages
  and Operating Systems}, ser. ASPLOS '18.\hskip 1em plus 0.5em minus
  0.4em\relax New York, NY, USA: ACM, 2018, pp. 622--636.

\bibitem{CheBoyer2009}
S.~Che, M.~Boyer, J.~Meng, D.~Tarjan, J.~W. Sheaffer, S.-H. Lee, and
  K.~Skadron, ``{Rodinia: A Benchmark Suite for Heterogeneous Computing},'' in
  \emph{{IEEE International Symposium on Workload Characterization}}, 2009.

\bibitem{KimBatten2014}
J.~Y. Kim and C.~Batten, ``{Accelerating Irregular Algorithms on GPGPUs Using
  Fine-Grain Hardware Worklists},'' in \emph{{47th Annual IEEE/ACM
  International Symposium on Microarchitecture}}, ser. {MICRO}, Dec 2014, pp.
  75--87.

\bibitem{BestaPodstawski2017}
M.~Besta, M.~Podstawski, L.~Groner, E.~Solomonik, and T.~Hoefler, ``{To Push or
  To Pull: On Reducing Communication and Synchronization in Graph
  Computations},'' in \emph{{Proceedings of the 26th International Symposium on
  High-Performance Parallel and Distributed Computing}}, ser. {HPDC}, 2017, pp.
  93--104.

\bibitem{SinclairAlsop2015}
M.~D. Sinclair, J.~Alsop, and S.~V. Adve, ``{Efficient GPU Synchronization
  without Scopes: Saying No to Complex Consistency Models},'' in
  \emph{{Proceedings of the 48th Annual IEEE/ACM International Symposium on
  Microarchitecture}}, ser. {MICRO}, December 2015.

\bibitem{SinclairAlsop2017}
------, ``{Chasing Away RAts: Semantics and Evaluation for Relaxed Atomics on
  Heterogeneous Systems},'' in \emph{{Proceedings of the 44th Annual
  International Symposium on Computer Architecture}}, ser. {ISCA}, 2017, pp.
  161--174.

\bibitem{AlsopSinclair2018}
J.~Alsop, M.~D. Sinclair, and S.~V. Adve, ``{Spandex: A Generalized Interface
  for Flexible Heterogeneous Coherence},'' in \emph{{Proceedings of the 45th
  International Symposium on Computer Architecture}}, ser. {ISCA}, June 2018.

\bibitem{HowerHechtman2014}
\BIBentryALTinterwordspacing
D.~R. Hower, B.~A. Hechtman, B.~M. Beckmann, B.~R. Gaster, M.~D. Hill, S.~K.
  Reinhardt, and D.~A. Wood, ``{Heterogeneous-Race-Free Memory Models},'' in
  \emph{{Proceedings of the 19th International Conference on Architectural
  Support for Programming Languages and Operating Systems}}, ser.
  {ASPLOS}.\hskip 1em plus 0.5em minus 0.4em\relax New York, NY, USA: ACM,
  2014, pp. 427--440. [Online]. Available:
  \url{http://doi.acm.org/10.1145/2541940.2541981}
\BIBentrySTDinterwordspacing

\bibitem{AlsopOrr2016}
J.~Alsop, M.~S. Orr, B.~M. Beckmann, and D.~A. Wood, ``{Lazy Release
  Consistency for GPUs},'' in \emph{{49th Annual IEEE/ACM International
  Symposium on Microarchitecture}}, ser. {MICRO}.\hskip 1em plus 0.5em minus
  0.4em\relax IEEE, 2016, pp. 1--14.

\bibitem{AdveHill1993}
S.~V. Adve and M.~D. Hill, ``{A Unified Formalization of Four Shared-Memory
  Models},'' \emph{TPDS}, pp. 613--624, June 1993.

\bibitem{DavisHu2011}
T.~A. Davis and Y.~Hu, ``{The University of Florida Sparse Matrix
  Collection},'' \emph{{ACM Transactions on Mathematical Software}}, vol.~38,
  no.~1, pp. 1:1--1:25, Dec. 2011.

\bibitem{CheBeckmann2013}
S.~Che, B.~M. Beckmann, S.~K. Reinhardt, and K.~Skadron, ``{Pannotia:
  Understanding Irregular GPGPU Graph Applications},'' in \emph{{IEEE
  International Symposium on Workload Characterization}}, ser. {IISWC}, Sept
  2013, pp. 185--195.

\bibitem{JaiganeshBurtscher2018}
J.~Jaiganesh and M.~Burtscher, ``A high-performance connected components
  implementation for gpus,'' in \emph{Proceedings of the 27th International
  Symposium on High-Performance Parallel and Distributed Computing}, ser. HPDC
  '18.\hskip 1em plus 0.5em minus 0.4em\relax ACM, 2018, pp. 92--104.

\bibitem{Li2019}
\BIBentryALTinterwordspacing
B.~Li, J.~Wei, J.~Sun, M.~Annavaram, and N.~S. Kim, ``An efficient gpu cache
  architecture for applications with irregular memory access patterns,''
  \emph{ACM Trans. Archit. Code Optim.}, vol.~16, no.~3, pp. 20:1--20:24, Jun.
  2019. [Online]. Available: \url{http://doi.acm.org/10.1145/3322127}
\BIBentrySTDinterwordspacing

\bibitem{Xu2019}
\BIBentryALTinterwordspacing
Z.~Xu, X.~Chen, J.~Shen, Y.~Zhang, C.~Chen, and C.~Yang, ``Gardenia: A graph
  processing benchmark suite for next-generation accelerators,'' \emph{J.
  Emerg. Technol. Comput. Syst.}, vol.~15, no.~1, pp. 9:1--9:13, Jan. 2019.
  [Online]. Available: \url{http://doi.acm.org/10.1145/3283450}
\BIBentrySTDinterwordspacing

\bibitem{MartinSorin2005}
M.~M.~K. Martin, D.~J. Sorin, B.~M. Beckmann, M.~R. Marty, M.~Xu, A.~R.
  Alameldeen, K.~E. Moore, M.~D. Hill, and D.~A. Wood, ``{Multifacet's General
  Execution-driven Multiprocessor Simulator (GEMS) Toolset},'' \emph{SIGARCH
  Computer Architecture News}, 2005.

\bibitem{AgarwalKrishna2009}
N.~Agarwal, T.~Krishna, L.-S. Peh, and N.~K. Jha, ``{GARNET: A Detailed On-chip
  Network Model Inside a Full-system Simulator},'' in \emph{{IEEE International
  Symposium on Performance Analysis of Systems and Software}}, 2009.

\bibitem{MagnussonChristensson2002}
P.~S. Magnusson, M.~Christensson, J.~Eskilson, D.~Forsgren, G.~Hallberg,
  J.~Hogberg, F.~Larsson, A.~Moestedt, and B.~Werner, ``Simics: A full system
  simulation platform,'' \emph{Computer}, vol.~35, no.~2, pp. 50--58, 2002.

\bibitem{BakhodaYuan2009}
A.~Bakhoda, G.~L. Yuan, W.~W.~L. Fung, H.~Wong, and T.~M. Aamodt, ``{Analyzing
  CUDA Workloads Using a Detailed GPU Simulator},'' in \emph{{IEEE
  International Symposium on Performance Analysis of Systems and Software}},
  2009.

\bibitem{AlsopSinclair2016}
J.~Alsop, M.~D. Sinclair, R.~Komuravelli, and S.~V. Adve, ``{GSI: A GPU Stall
  Inspector to Characterize the Sources of Memory Stalls for Tightly Coupled
  GPUs},'' in \emph{{IEEE International Symposium on Performance Analysis of
  Systems and Software}}, ser. {ISPASS}, April 2016, pp. 172--182.

\bibitem{NasreBurtscher2013}
R.~Nasre, M.~Burtscher, and K.~Pingali, ``{Atomic-free Irregular Computations
  on GPUs},'' in \emph{{Proceedings of the 6th Workshop on General Purpose
  Processor Using Graphics Processing Units}}, ser. {GPGPU-6}, 2013, pp.
  96--107.

\bibitem{ElteirLin2011}
M.~Elteir, H.~Lin, and W.~chun Feng, ``Performance characterization and
  optimization of atomic operations on amd gpus,'' \emph{2011 IEEE
  International Conference on Cluster Computing}, pp. 234--243, 2011.

\bibitem{XiaoFeng2010}
S.~Xiao and W.~chun Feng, ``Inter-block gpu communication via fast barrier
  synchronization,'' \emph{2010 IEEE International Symposium on Parallel \&
  Distributed Processing (IPDPS)}, pp. 1--12, 2010.

\bibitem{KaleemVenkat2016}
R.~Kaleem, A.~Venkat, S.~Pai, M.~Hall, and K.~Pingali, ``{Synchronization
  Trade-Offs in GPU Implementations of Graph Algorithms},'' in \emph{{IEEE
  International Parallel and Distributed Processing Symposium}}, ser. {IPDPS},
  May 2016, pp. 514--523.

\bibitem{OrrChe2015}
M.~S. Orr, S.~Che, A.~Yilmazer, B.~M. Beckmann, M.~D. Hill, and D.~A. Wood,
  ``{Synchronization Using Remote-Scope Promotion},'' in \emph{{Proceedings of
  the Twentieth International Conference on Architectural Support for
  Programming Languages and Operating Systems}}, ser. {ASPLOS}.\hskip 1em plus
  0.5em minus 0.4em\relax New York, NY, USA: ACM, 2015.

\bibitem{HamWu2016}
T.~J. Ham, L.~Wu, N.~Sundaram, N.~Satish, and M.~Martonosi, ``Graphicionado: A
  high-performance and energy-efficient accelerator for graph analytics,'' in
  \emph{The 49th Annual IEEE/ACM International Symposium on Microarchitecture},
  ser. MICRO-49.\hskip 1em plus 0.5em minus 0.4em\relax Piscataway, NJ, USA:
  IEEE Press, 2016, pp. 56:1--56:13.

\bibitem{PrabhakarZhang2017}
R.~Prabhakar, Y.~Zhang, D.~Koeplinger, M.~Feldman, T.~Zhao, S.~Hadjis,
  A.~Pedram, C.~Kozyrakis, and K.~Olukotun, ``Plasticine: A reconfigurable
  architecture for parallel patterns,'' in \emph{2017 ACM/IEEE 44th Annual
  International Symposium on Computer Architecture (ISCA)}, June 2017, pp.
  389--402.

\bibitem{AhmadKhan2016}
M.~Ahmad and O.~Khan, ``{GPU Concurrency Choices in Graph Analytics},'' in
  \emph{{IEEE International Symposium on Workload Characterization}}, ser.
  {IISWC}, Sept 2016, pp. 1--10.

\bibitem{HongOguntebi2011}
S.~Hong, T.~Oguntebi, and K.~Olukotun, ``Efficient parallel graph exploration
  on multi-core cpu and gpu,'' in \emph{Proceedings of the 2011 International
  Conference on Parallel Architectures and Compilation Techniques}, ser. PACT
  '11.\hskip 1em plus 0.5em minus 0.4em\relax Washington, DC, USA: IEEE
  Computer Society, 2011, pp. 78--88.

\bibitem{LiBecchi2013}
D.~Li and M.~Becchi, ``Deploying graph algorithms on gpus: An adaptive
  solution,'' in \emph{2013 IEEE 27th International Symposium on Parallel and
  Distributed Processing}, May 2013, pp. 1013--1024.

\bibitem{HestnessKeckler2015}
J.~Hestness, S.~W. Keckler, and D.~A. Wood, ``{GPU Computing Pipeline
  Inefficiencies and Optimization Opportunities in Heterogeneous CPU-GPU
  Processors},'' in \emph{{2015 IEEE International Symposium on Workload
  Characterization}}, ser. {IISWC}, October 2015.

\bibitem{XuJeon2014}
Q.~Xu, H.~Jeon, and M.~Annavaram, ``{Graph Processing on GPUs: Where are the
  Bottlenecks?}'' in \emph{{2014 IEEE International Symposium on Workload
  Characterization}}, ser. {IISWC}, Oct 2014, pp. 140--149.

\bibitem{EyermanHeirman2018}
S.~Eyerman, W.~Heirman, K.~D. Bois, J.~B. Fryman, and I.~Hur, ``Many-core graph
  workload analysis,'' \emph{2018 International Conference for High Performance
  Computing, Networking, Storage and Analysis (SC)}, 2018.

\bibitem{Ahmad2019}
M.~{Ahmad}, H.~{Dogan}, C.~J. {Michael}, and O.~{Khan}, ``Heteromap: A runtime
  performance predictor for efficient processing of graph analytics on
  heterogeneous multi-accelerators,'' in \emph{2019 IEEE International
  Symposium on Performance Analysis of Systems and Software (ISPASS)}, March
  2019, pp. 268--281.

\bibitem{Sorensen2019}
T.~Sorensen, S.~Pai, and A.~F. Donaldson, ``One size doesn’t fit all:
  Quantifying performance portability of graph applications on gpus,'' in
  \emph{2019 IEEE International Symposium on Workload Characterization
  (IISWC'19)}, 11 2019.

\end{thebibliography}

\end{document}